\documentclass[11pt]{article}
\usepackage{multirow}
\usepackage{appendix}
\usepackage{graphicx} 
\usepackage{siunitx} 
\usepackage{booktabs} 
\usepackage{subfiles}
\usepackage{tabularx}
\usepackage{changepage}
\usepackage{longtable}
\usepackage{adjustbox}
\usepackage{lscape }
\usepackage{pdflscape }
\usepackage{amsmath}
\usepackage{setspace}
\usepackage[bottom]{footmisc}
\usepackage{caption}
\usepackage{subcaption}
\usepackage{xcolor}
\usepackage[ pdftex, plainpages = false, pdfpagelabels, 
pdfpagelayout = useoutlines,
bookmarks,
bookmarksopen = true,
bookmarksnumbered = true,
colorlinks=true, 
linktocpage=true,
citecolor = blue,
urlcolor= blue,
linkcolor = red]{hyperref} 
\usepackage{graphicx,caption}
\usepackage{changepage}

\usepackage{rotating}
\usepackage[sort]{natbib}
\usepackage{shortvrb}
\usepackage{eurosym}
\usepackage{float}
\usepackage{placeins}
\usepackage{lipsum} 
\usepackage[paperwidth=210mm, paperheight=297mm, top=2cm, margin=2cm, centering]{geometry}
\usepackage[labelfont=bf]{caption}
\usepackage{amsmath}
\usepackage[USenglish]{babel}
\usepackage{bbding}

\usepackage{sectsty}
\usepackage{tocloft}
\newcolumntype{L}[1]{>{\raggedright\arraybackslash}p{#1}}
\newcolumntype{C}[1]{>{\centering\arraybackslash}p{#1}}
\newcolumntype{R}[1]{>{\raggedleft\arraybackslash}p{#1}}

\usepackage{afterpage}

\usepackage{tikz}
\usetikzlibrary{shapes,arrows,positioning}
\usetikzlibrary{automata,arrows,positioning,calc}
\usepackage{authblk}

\usepackage{pgfplots}% lädt auch tikz
\pgfplotsset{compat=newest}
\usetikzlibrary{decorations.pathreplacing,calc}

\usepackage{multirow}
\usepackage{appendix}
\usepackage{graphicx} 
\usepackage{siunitx} 
\usepackage{booktabs} 
\usepackage{subfiles}
\usepackage{tabularx}
\usepackage{changepage}
\usepackage{longtable}
\usepackage{adjustbox}
\usepackage{lscape }
\usepackage{pdflscape }
\usepackage{amsmath}
\usepackage{setspace}
\usepackage{multibib}
\usepackage{rotating}
\usepackage{natbib}
\usepackage{shortvrb}
\usepackage{eurosym}
\usepackage{float}
\usepackage{placeins}
\usepackage{lipsum} 
\usepackage{tikz}
\usetikzlibrary{shapes,arrows,positioning}
\usetikzlibrary{automata,arrows,positioning,calc}
\usepackage{authblk}
\usepackage{xurl}
\usepackage{tocloft}

\usepackage[USenglish]{babel}

\usepackage{pgfplots}% lädt auch tikz
\pgfplotsset{compat=newest}
\usetikzlibrary{decorations.pathreplacing,calc}

\usepackage{natbib}
\usepackage{sectsty}
\usepackage{tocloft}
\usepackage{afterpage}

\usepackage{titlesec}

\renewenvironment{abstract}
{
	\begin{center}
		\bfseries \abstractname\vspace{-0.5em}\vspace{0pt}
	\end{center}
	\list{}{%
		\setlength{\leftmargin}{0mm}% <---------- CHANGE HERE
		\setlength{\rightmargin}{0mm}%
	}%
	\item\relax}
{\endlist}

\usepackage[USenglish]{babel}
\usepackage[T1]{fontenc}
\usepackage{tgpagella}

\title {\textbf{Beyond the heat: \\
		 The mental health toll of temperature and humidity in India}}

\author[1,2]{Manuela Fritz\footnote{Corresponding author: Manuela Fritz, TUM School of Social Sciences and Technology. Richard Wagner-Straße 1, 80333 Munich, Germany. Mail: manuela.fritz@tum.de.}}

\affil[1]{\small{\textit{TUM School of Social Science and Technology, Professorship for Global Health, Munich, Germany}}}
\affil[2]{\small{\textit{University of Passau, Chair of Development Economics, Passau, Germany}}}
\date{}

\begin{document}
\begin{titlepage}
\onehalfspacing
	\clearpage\maketitle
	\thispagestyle{empty}

	\begin{abstract}
		\noindent	
\vspace{-1cm}
	\singlespacing
Evidence on the heat-mental health nexus remains mixed. I show that this can be partly explained by previous studies focusing solely on temperature while neglecting temperature-humidity interactions. Using a measure that considers both indicators (wet bulb temperature), I assess the causal link between extreme heat and mental health, and its heterogeneity across socioeconomic indicators. I combine self-reported depression and anxiety levels from three Indian WHO-SAGE survey waves with climate data, leveraging quasi-random variation in heat exposure due to survey timing and location. The results reveal that extreme heat increases the risk of depression but not of anxiety. Importantly, these effects are consistently smaller when humidity is not considered. Finally, the study provides evidence that the District Mental Health Program plays a protective role in mitigating adverse mental health effects. The findings suggest that the costs induced by climate change need to account for the economic consequences of deteriorated mental health.
\\
	\end{abstract}

\noindent
\textbf{Keywords:} \\
\noindent
 Mental health, depression, heatwaves, wet bulb temperature, socioeconomic heterogeneity, India, \\
 District Mental Health Program. \\

\noindent
\textbf{JEL Codes:}  \\
\noindent
 I1, I15, I31, Q54.  \\
\vspace{3cm}	

\noindent
\textbf{Acknowledgments:}  \\
\noindent
I gratefully acknowledge funding from the Deutsche Forschungsgemeinschaft (DFG, German Research Foundation) under the Project Number 537845359. \\
\vspace{1cm}	

\end{titlepage}

%%%%%%%%%%%%%%%%%%%%%%%%%%%%%%%%%%%%%%%%%%%%%%%%%%%%%%%
\parskip 5pt
\onehalfspacing

\section{Introduction}

The scientific discourse on the consequences of climate change has spurred a substantial body of literature examining various economic, societal and psychological outcomes influenced by heat.\footnote{Recent research has explored the impact of heat on worker productivity and aggregate GDP \citep{dell2012temperature, lopalo2023temperature, heyes2022hot}, migration \citep{cattaneo2016migration}, judicial decisions \citep{heyes2019temperature}, violence and conflict \citep{hsiang2013quantifying,sanz2018heat}, and cognitive performance \citep{zivin2020temperature}.} Nonetheless, the primary direct effects of extreme heat remain health-related outcomes. Extensive research has documented the detrimental impact of heat on mortality \citep{ballester2023heat, barreca2016adapting, deschenes2011climate, burgess2017weather} and morbidity \citep{fritz2022temperature, karlsson2018population, white2017dynamic}. However, the relationship between heat and mental health outcomes, such as suicide rates, depressive symptoms, and emotional well-being, has received comparatively less attention, with most existing evidence derived from the US or European countries \citep{berry2018case,mullins2019temperature, noelke2016increasing,obradovich2018empirical}. In contrast, the evidence for the link between heat exposure and mental health in low- and middle-income countries (LMICs) is still limited \citep{thompson2018associations, liu2021there}. 

Heat effects on mental health outcomes in LMICs could deviate from effects found for Western or high income countries for several reasons. Most of LMICs lie in the Global South and their populations are exposed not only to on average hotter, but also more humid climatic conditions than individuals in the US or European countries. Moreover, a larger share of the population works in agriculture or other physically very demanding jobs, often to a large extent outdoors, have less access to air conditioning, and are hence more exposed to heat extremes. Further, given that a large share of the world's poor live in the Global South and poverty has been shown to affect mental resources \citep{mani2013poverty, ridley2020poverty}, the link between heat and mental health could be reinforced by poverty, which itself can be deepened through heatwaves if they are accompanied by droughts causing harvest failures. Lastly, while high-income countries have seen a decrease in their mental disorder disease burden in the last decade (in terms of disability adjusted life years (DALYs)), the burden in LMICs has been steadily rising \citep{gbd2019}. In India, the context of this study, the proportion of the disease burden due to mental disorders has doubled since 1990, with nowadays one in seven Indians suffering from mental illnesses \citep{sagar2020burden}. Yet, at the same time, treatment and care for mental health remains limited, with a treatment gap of 80\% for common mental disorders \citep{jayasankar2022epidemiology}.

India and more broadly the South Asian region are among the areas for which the consequences of climate change are predicted to be most severe. Specifically, according to the Intergovernmental Panel of Climate Change (IPCC), the region belongs to those in which the combination of high temperature and high humidity levels will lead to heat exposure that induces extreme risks to human health. Similarly, \cite{lopalo2019quantifying} argue that the predicted loss in GDP due to health and productivity losses driven by climate change in India has been substantially underestimated since humidity was not accounted for. Yet, so far, the interaction between high temperatures and high humidity levels has not received particular attention in the literature assessing the link between heat and (mental) health in general and in India in particular; previous studies either did not include humidity levels at all \citep[e.g.][]{burgess2017weather} or only as control variable \citep[e.g.][]{carleton2017crop, pailler2018effects}. 

Moreover, the evidence for the heat-mental health relationship for LMICs so far focuses primarily on the effects of extreme heat on suicides rates and hospital admissions, mostly driven by exacerbation of already existing mental illnesses. For this link, the literature finds robust evidence for a relationship between extreme temperatures and exacerbated illness. For example, increased suicide rates due to increased temperatures have been documented in India \citep{carleton2017crop},  Mexico \citep{burke2018higher}, or China \citep{luan2019associations}, and increases in mental health related hospital admissions have been recorded in Vietnam \citep{trang2016heatwaves} and China \citep{liu2019influence, wang2018effect}. The literature is much more limited and inconclusive when it comes to self-reported mental health states. Understanding the effects of heat on such ``lower level'' mental health outcomes is important since they will most likely affect a much larger share of the population (also those without prior existing mental health illnesses). While \cite{pailler2018effects} in the context of India find that temperature (and rainfall) variability reduces the self-reported ability to cope with one's life, they do not find any effects on key depressive symptoms. Similarly, \cite{xue2019declines} do not find a significant relationship between mean temperature levels and self-rated mental health scores in China. In contrast, \cite{zhang2023heatwave} show that heatwaves in East China reduce self-reported happiness levels. One factor explaining these mixed results could be that mental disorders remain stigmatized, hence might be under-reported in interviews and might not be captured in an analysis based on survey responses \citep{bharadwaj2017mental}. This is also true in the Indian context \citep{kaur2021systematic}. 

Another reason might be that simply focusing on temperature and ignoring humidity levels has led to an underestimation of the actual effects; a channel that I explore in this study. Specifically, I argue that previous studies have underestimated the heat-mental health nexus by only focusing on temperature and neglecting the interaction with high humidity levels. I address this shortcoming by using a temperature measure that incorporates the interaction between heat and humidity, namely the wet bulb temperature, which provides a more accurate prediction of actual heat stress \citep{saeed2021deadly,matthews2017communicating}, and show that there is a robust link between extreme wet bulb temperatures and and adverse mental health outcomes measured as self-reported depressive symptoms. 

For identification, I rely on the temporal and geographical variation in the roll-out of the Indian WHO-SAGE surveys. Under the assumption that the survey roll-out was independent of the weather conditions (which is plausible, given the strict time protocol such large-scale studies follow), the variation in the respondents' exposure to the frequency and severity of heat events in the 30 days preceding the survey interview can be considered as good as random. I make use of a ``temperature bin regression'' approach, as it has become standard in the literature \citep[e.g.][]{deschenes2011climate, deschenes2009climate, white2017dynamic, karlsson2018population}, but also use multiple other heat indicators to capture effects at both the extensive and intensive margin.

This study most closely links to the studies of \cite{carleton2017crop} and \cite{pailler2018effects}, showing for the context of India that heat increases suicides and decreases self-reported mental well-being. Both studies, however, do not consider humidity levels and focus on the longer-term effects of heat events in the previous agricultural season on mental health. In contrast, I rely on exposure to recent heat events over a one-month horizon and consider heat and humidity jointly, which allows me to focus on short-term physiological mechanisms as a plausible explanation between heat and mental health outcomes. I find that a heatwave in the 30 days preceding the survey increases the probability of suffering from severe or extreme depression by 24\% at the extensive margin (exposure to a heatwave lasting two or more days) and by 6\% at the intensive margin (one additional heat day). I find no significant effects on anxiety levels. Crucially, these effects for depression become substantially smaller in size and significance or disappear completely when only temperature levels are used as the explanatory variable, without accounting for the temperature-humidity interaction. 

I then investigate whether an existing mental health intervention in India -- the District Mental Health Program -- has the potential to address some of the negative mental health impacts of heat. Making use of the staggered roll-out of this program across the country, I find that the effect size of the heat-mental health relationship is significantly smaller in districts that had access to the program, suggesting a protective role of the program in mitigating adverse mental health effects.  

The remainder of this paper is organized as follows. In Section 2, I provide details about the different data sources. Section 3 presents the empirical strategy. This is followed by the presentation of the results in Section 4, including robustness checks, an investigation of the time dynamics, and heterogeneity assessment across a range of socioeconomic indicators. In Section 5, I discuss potential channels through which heat affects mental health outcomes in the given context. Section 6 investigates the effectiveness of the District Mental Health Program for alleviating the heat-mental health link. In Section 7, I discuss the results and their implications and conclude. 

\section{Data}
To identify the impact of high wet bulb temperatures on mental health outcome, I use individual level survey data from three cross-sectional waves of the Indian WHO Study on Global Ageing and Adult Health (WHO-SAGE)\footnote{A part of the WHO-SAGE study was designed as longitudinal study with panel individuals. However, the publicly available data do not contain an individual identifier that would allow to link individuals across survey waves. This was also confirmed through e-mail correspondence with the Indian International Institute for Population Sciences (IIPS). Therefore, I employ only the cross-sectional dimension. The IIPS is currently working on the provision of necessary information to build a panel data set, but this was not yet available at the time this study was conducted. Also, a third follow-up has been conducted in 2018/2019, but data are not yet publicly available.} and merge them with data on the weather conditions during the 30 days preceding the interview date. 

\subsection{Mental health outcomes: WHO-SAGE study}
The Indian WHO-SAGE study is a nationally representative household survey with the objective to collect information on adult health and well-being with a focus on the population above the age of 50 \citep{sage1, sage2}. Additionally, comparison samples for individuals aged 18-49 were included in the survey. The baseline survey (Wave 0) was collected as part of the WHO World Health Survey (WHS, in 2002/2003) and served as sampling frame for two follow-up surveys in 2007 (Wave 1) and 2015 (Wave 2). I employ all three cross-sectional waves. Geo-coded locations of the primary sampling units (PSUs, which are villages in rural areas and wards in urban areas) are drawn from the WHS and are merged to the PSUs in  follow-up Waves 1 and 2. For two PSUs, no geo-locations were available in Wave 0 and four new PSUs were added in Wave 1, yet without available geo-coded locations, resulting in a sample of 371 PSUs with geo-locations observed in all three waves.  

All three survey waves were conducted in six Indian states (Assam, Karnataka, Maharashtra, Rajasthan, Uttar Pradesh and West Bengal), which together account for almost 50\% of the Indian population. In Wave 0, around 10,300 individuals were interviewed, with follow-up sample sizes of around 11,000 in Wave 1 and around 9,000 in Wave 2. Face-to-face interviews were used to obtain information about the individuals' subjective and objective health status, including health state descriptions, presence of chronic conditions, subjective well-being and quality of life, access to health care and health care utilization. The study further included physical and cognition tests, such as vision tests, blood pressure, and verbal and digit recall tests. Additionally, individual and household information about demographic and socioeconomic characteristics, such as age, gender, marital status, assets, income and expenditures were collected. 

The main variables of interest are the self-reported mental health states. Specifically, I make use of indicators of depressive and anxiety symptoms during the last 30 days. Respondents were asked ``Overall in the last 30 days, how much of a problem did you have with feeling sad, low, or depressed'' and ``Overall in the last 30 days, how much of a problem did you have with worry or anxiety''. Answers could be in the range of 1 (none) and 5 (extreme). For both health states, I define a binary indicator equal to one if individuals reported to suffer from severe or extreme depression or from severe or extreme anxiety (scale units 4 and 5).

Table \ref{sumstats}, Panel 1, presents an overview of the characteristics of the respondents, pooled over all three survey waves. The sample consists of slightly more women than men. The mean age is 48 years, with a standard deviation of 17 years. Respondents have on average 4.6 years of education, 77\% are married and 75\% live in a rural location. Seventy-five percent have access to electricity and about 43\% suffer from a chronic health condition.\footnote{The following chronic conditions are included: arthritis, stroke, angina, diabetes, chronic lung diseases, and hypertension.}

The mean depression score is 1.82, which is lower than the mean anxiety score of 2.01, indicating that respondents suffered more from worry/anxiety than from depression. Similarly, 8\% reported to have suffered from severe or extreme depression in the last 30 days, whereas 13\% reported to have suffered from extreme or severe worry/anxiety in the last 30 days.

\subsection{Weather, climate and environmental data}
\subsubsection{Wet bulb temperature}
The information about the exact date of the survey interview together with the geo-locations of the PSUs allows to match the survey data with meteorological data on the temperature and humidity conditions during the 30 days preceding the survey. This paper focuses on wet bulb temperature as opposed to dry bulb temperature (the ``conventional'' air temperature), with the wet bulb temperature being a non-linear function of dry bulb temperature, moderated by relative humidity levels. At 100\% humidity, dry bulb and wet bulb temperature are identical, whereas wet bulb temperature readings are always lower than dry bulb temperature if humidity levels are below 100\%. Wet bulb temperature is the lowest temperature that can be reached through cooling by the evaporation of water. Intuitively, wet bulb temperature is the temperature that will be reached if a thermometer is covered in wet cloth. The higher the water saturation in the air, the less water can evaporate and the less cooling is possible. For humans, this means that the cooling process through sweating is less effective the higher the humidity level at a given temperature. A wet bulb temperature of 35° (corresponding for example to a dry bulb temperature of 45°C at 50\% relative humidity or 40°C at 75\% humidity) is considered the limit in which humans could no longer survive for a sustained period of time \citep{raymond2020emergence}.

The information on the wet bulb temperature are extracted from the NASA GLDAS Noah Land Surface Model \citep{NASA1}, which is a gridded data set with a spatial resolution of 0.25°$\times$0.25° and a 3-hourly temporal resolution. While this data set does not contain the wet bulb temperature directly, I use the information on dry bulb temperature, air pressure, and specific humidity to calculate wet bulb temperature, as specified in Appendix A. The 3-hourly observation are aggregated to daily averages by taking the mean of all eight observations per 24-hour interval. The resulting daily temperature grids are merged to the respective PSU by using inverse-distance weighting of the four nearest grid points to each PSU. I also construct the long-run average wet bulb temperature and its standard deviation for a given day over the years 2000-2016, which are used to control for long-term climate conditions.\footnote{To construct the long-run daily wet bulb temperature, I use the monthly observations from the GLDAS Noah Land Surface Model over the period 2000-2016, calculate the long-run monthly means and weigh them by the number of days in the respective month in the last 30 days.}

\begin{figure}[!h]
	\includegraphics[scale=.15]{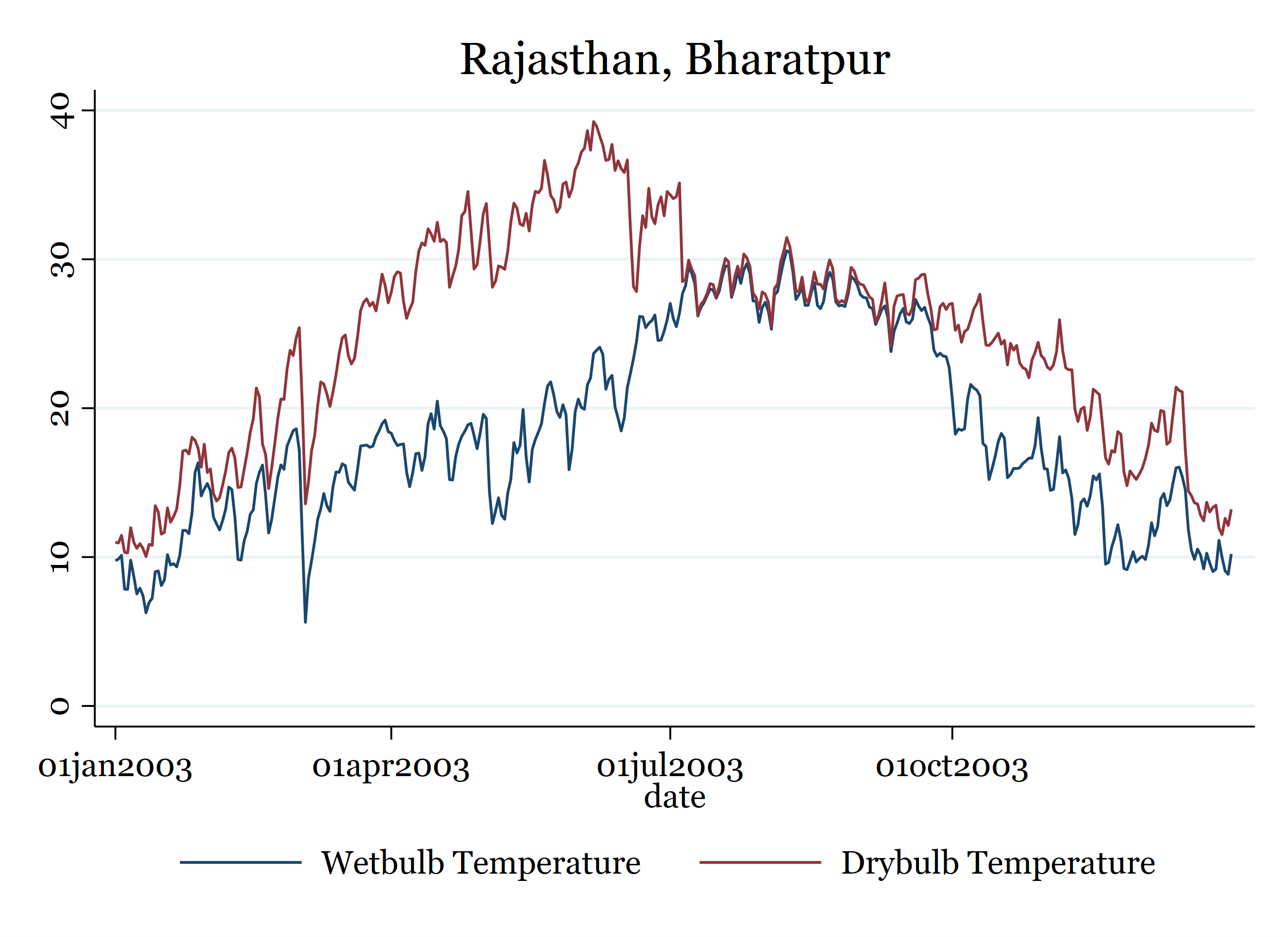}
	\centering
	\caption[Wet bulb and dry bulb temperature in 2003 in Rajasthan]{\textbf{Wet bulb and dry bulb temperature in 2003 in Rajasthan}}
	\label{trend_wetbulb}
\end{figure}

Figure \ref{trend_wetbulb} exemplary presents the 2003 wet bulb and dry bulb temperature trend in Bharatpur in the state Rajastan. It shows that not only the levels are different (as explained above), but that also the trend in the two temperature measures occasionally differs significantly. In particular, while the highest dry bulb temperatures are reached in June, the highest wet bulb temperatures are only reached in August and September, when humidity levels are highest (indicated by almost identical dry bulb and wet bulb temperatures).

\subsubsection{Control variables}
From the same NASA data set and using the same approach as above, I obtain information on average wind speed. I further control for rainfall levels and air pollution, to be able to identify the pure heat effect. Data on daily rainfall levels are obtained from the Indian Meteorological Department (IMD), which similarly provides the data in the form of a gridded data set with a spatial resolution of 0.25°$\times$0.25° \citep{pai2014development}. Information on air pollution is drawn from the NASA MERRA-2 data set, which is available on a 0.5°$\times$0.625° grid and with an hourly time resolution \citep{MERRA2}. These hourly observations are averaged to daily observations and surface particulate matter concentration (PM2.5) is calculated by using the formula displayed in Appendix A. The assignment of daily grids to the respective PSU is analogous to the inverse distance weighting procedure described above.
   
Lastly, I obtain information about the Koeppen-Geiger climate classifications from \cite{beck2018present}. Koeppen-Geiger classifications specify different global climate regions according to their seasonal temperature and rainfall patterns. The employed data set provides for each 0.1°$\times$0.1° degree grid of the world the respective climate classifications for the period 1990-2010. I use this high resolution data set to assign each PSU its corresponding climate classification by using the Koeppen-Geiger climate classifications of the nearest grid point (the average distance between the coordinates of the PSU and the nearest grid point is 0.65km). 

\begin{figure}[!h]
	\includegraphics[scale=.4]{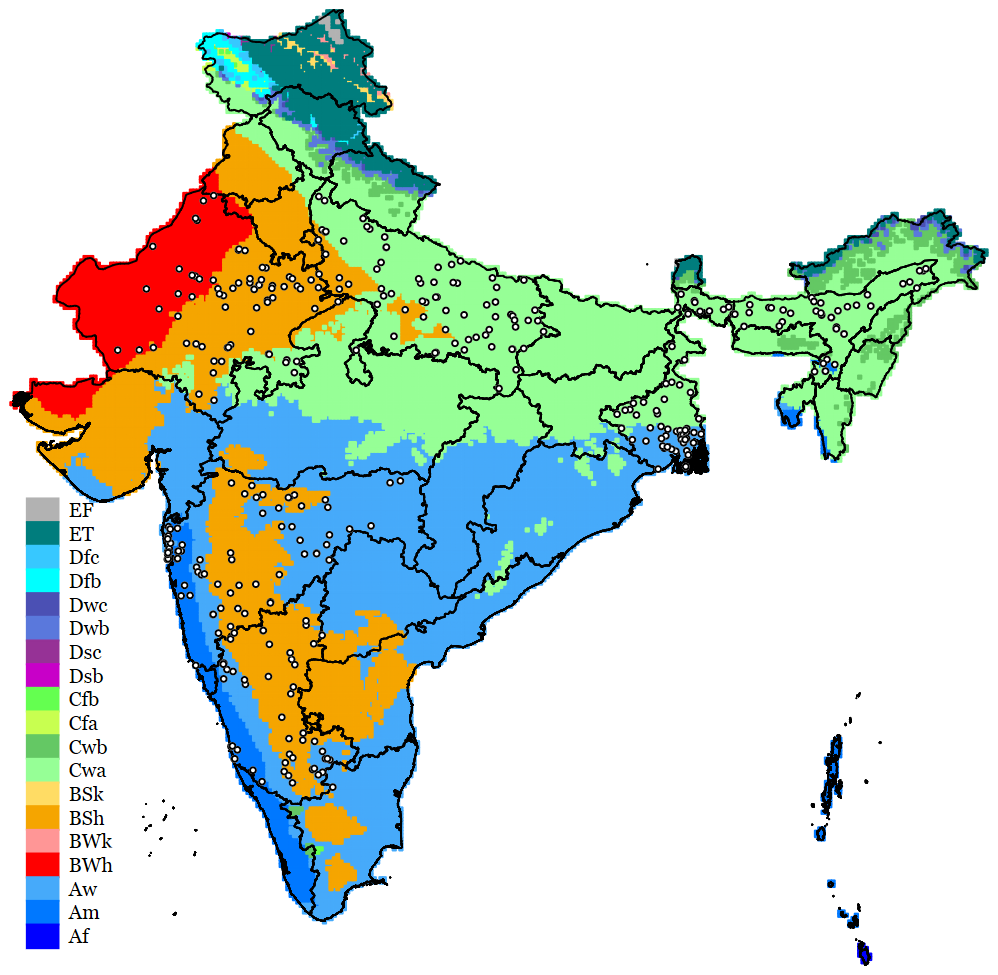}
	\centering
	\caption[Koeppen-Geiger Climate classifications and distribution of sampling clusters]{\textbf{Koeppen-Geiger Climate classifications and distribution of sampling clusters -} The Koeppen-Geiger climate classifications in which the sampling clusters fall are: \textit{Am}: tropical -- monsoon, \textit{Aw}: tropical -- savanna, dry winter, \textit{BWh}: dry -- arid desert -- hot, \textit{BSh}: dry -- semi-arid steppe -- hot, \textit{Cwa}: temperate -- dry winter -- hot summer, \textit{Cwb}; temperate -- dry winter -- warm summer.}
	\label{map}
\end{figure}

Figure \ref{map} presents the Koeppen-Geiger climate zones of India together with the  geographical distribution of the PSUs.  Note that climate zones are not homogeneous within states (or even district) borders. Therefore, I control for climate region fixed-effects in the empirical specification below (in addition to state fixed-effects), resulting in a comparison of individuals that are on average exposed to the same climate conditions, but differ in the exposure to heat during the 30 days preceding the survey.

Table \ref{sumstats}, Panel 2, presents the summary statistics of the temperature and climate variables (summary statistics for dry bulb temperatures are presented in Appendix B). The average daily \textit{mean} wet bulb temperature is about 22°C and reaches maximum values of more than 28°C (the highest daily \textit{maximum} wet bulb temperature observed in the sample is 32.2°C). Average rainfall levels in the last 30 days range between 0 and 1355 mm and deviate between -740 mm and + 766 mm from the long-run 10 year average monthly rainfall. Mean daily levels of air pollution over the last 30 days are very high by international standards, with a mean of 34 \textmu g/m\textsuperscript{3}.

\begin{table}[]
		\caption{\label{sumstats}Summary statistics}
		\centering
		\small
\begin{tabular}{lrrrrr}
		\midrule
		\midrule 
		& N    & Mean   & SD     & Min   & Max     \\
			\midrule
\textbf{Panel 1: Individual level variables}                            &       &        &        &         &         \\
Depression (scale)                                      & 29,541 &	1.82& 	0.99	& 1	&5      \\
Severe or   extreme depression (1=yes)                & 29,541 & 0.08   & 0.28   & 0       & 1       \\
Worry/Anxiety (scale)                                & 29,494 & 2.01   & 1.08   & 1       & 5       \\
Severe or extreme   anxiety (1=yes)                   & 29,494 & 0.13   & 0.33   & 0       & 1       \\
Male (1=yes)                                          & 29,719&	0.45&	0.50&	0	&1  \\
Age                                                   & 29,718&	48.36	&17.13 &	15&	106    \\
Years of   education                                  & 28,124 & 4.61   & 5.07   & 0       & 30      \\
Married (1=yes)                                       & 29,718 & 0.77   & 0.42   & 0       & 1       \\
Chronic   condition (1=yes)                           & 29,514 & 0.43   & 0.49   & 0       & 1       \\
Ever   diagnosed with depression (1=yes)              & 29,299 & 0.06   & 0.25   & 0       & 1       \\
Rural (1=yes)                                         & 30,300 & 0.75   & 0.43   & 0       & 1       \\
Access to   electricity (1=yes)                       & 29,767 & 0.75   & 0.43   & 0       & 1       \\
Household owns   a bike (1=yes)                       & 30,177 & 0.61   & 0.49   & 0       & 1       \\
Household   owns a car (1=yes)                        & 29,058 & 0.11   & 0.31   & 0       & 1       \\
&       &        &        &         &         \\
\textbf{Panel 2: Temperature and climate variables (N=30,300) }         &       &        &        &         &         \\
Mean wet bulb   temperature (°C)                      &       & 21.86  & 3.99   & 3.62    & 28.37   \\
&       &        &        &         &         \\
\textit{Temperature bins (distribution in last 30 days) }              &       &        &        &         &         \\
\# Days \textless 16.5°C                                      &       & 5.08   & 8.39   & 0       & 30      \\
\# Days 16.5°C-18°C                                           &       & 2.92   & 3.88   & 0       & 18      \\
\# Days 18°C-19.5°C                                           &       & 2.74   & 3.64   & 0       & 21      \\
\# Days 19.5°C-21°C                                           &       & 3.03   & 4.12   & 0       & 27      \\
\# Days 21°C-22.5°C                                           &       & 3.77   & 5.33   & 0       & 29      \\
\# Days 22.5°C-24°C                                           &       & 3.52   & 5.02   & 0       & 29      \\
\# Days 24°C-25.5°C                                           &       & 4.13   & 6.51   & 0       & 30      \\
\# Days 25.5°C-27°C                                           &       & 3.77   & 6.80   & 0       & 28      \\
\# Days  \textgreater 27°C                                     &       & 1.01   & 3.26   & 0       & 23      \\
&       &        &        &         &         \\
\# Consecutive days  \textgreater 27°C                 &       & 0.57    & 1.69  & 0       & 11       \\
Heatwave ($\geq$   2 days $\geq$ 27°)                 &       & 0.12    & 0.32   & 0       & 1       \\
Heatwave ($\geq$   3 days $\geq$ 27°)                 &       & 0.08    & 0.28   & 0       & 1       \\
Heatwave ($\geq$   2 days $\geq$ 28°)                 &       & 0.02    & 0.14   & 0       & 1       \\
&       &        &        &         &         \\
&       &        &        &         &         \\
Rainfall,   last 30 days (mm)                         &       & 111.71 & 184.94 & 0       & 1355.76 \\
Rainfall deviation   from long-run mean, last 30 days (mm) &       & 6.04   & 88.17  & -746.14 & 766.25  \\
Average wind   speed, last 30 days (m/s)                   &       & 2.97   & 1.06   & 0.93     & 7.50     \\
Average air   pollution (PM2.5), last 30 days (\textmu g/m\textsuperscript{3})         &       & 34.67  & 11.28  & 7.03    & 72.57   \\
&       &        &        &         &         \\
\textit{Koeppen-Geiger climate classifications }               &       &        &        &         &         \\
Am                                                    &       & 0.09   & 0.29   & 0       & 1       \\
Aw                                                    &       & 0.22   & 0.42   & 0       & 1       \\
BWh                                                   &       & 0.05   & 0.22   & 0       & 1       \\
BSh                                                   &       & 0.28   & 0.45   & 0       & 1       \\
Cwa                                                   &       & 0.35   & 0.48   & 0       & 1       \\
Cwb                                                   &       & 0.01   & 0.07   & 0       & 1       \\
&       &        &        &         &         \\
\textbf{Survey waves     }                                     &       &        &        &         &         \\
Survey wave 0, year 2003                                      &       & 0.34   & 0.47   & 0       & 1       \\
Survey wave 1, year 2007                                      &       & 0.37   & 0.48   & 0       & 1       \\
Survey wave 2, year 2015                                      &       & 0.29   & 0.46   & 0       & 1     \\

		\midrule   
		\midrule 
		\multicolumn{6}{p{16cm}}{\textit{Notes:} The Koeppen-Geiger climate classification system describes the following climate regions: \textit{Am}: tropical -- monsoon, \textit{Aw}: tropical -- savanna, dry winter, \textit{BWh}: dry -- arid desert -- hot, \textit{BSh}: dry -- semi-arid steppe -- hot, \textit{Cwa}: temperate -- dry winter -- hot summer, \textit{Cwb}; temperate -- dry winter -- warm summer.}   

	\end{tabular}
\end{table}

\section{Empirical strategy}
Identification comes from the variation in exposure to different wet bulb temperatures in the 30 days preceding the survey conditional on a series of geographical and time fixed-effects. I make use of several different empirical specifications to capture the effects of wet bulb heat on mental health at the extensive and intensive margin. I start with an empirical specification which has become standard in the climate literature \citep[e.g.][]{deschenes2009climate, karlsson2018population, burgess2017weather}. Specifically, I use a non-parametric approach, for which I define nine 1.5°C wet bulb temperature bins and use the number of days that fell in each of these bins in the 30 days prior of the survey interview. This allows to estimate the effect of one additional extremely hot day relative to a day in a given reference category on the state of mental health over the last 30 days. The top bin (>27°C) is chosen to correspond to the 95\% percentile of the mean wet bulb temperature distribution over the three survey waves. Within the sample, on average one day over the last 30 days reached such temperatures (see Table \ref{sumstats}). The empirical specification reads

\begin{equation}\label{equa1}
Mental_{idc} = \sum_{t=1,t \neq 5}^{9} \alpha_t Wetbulb_{dc}^{t} + \beta \mathbf{X}_{idc} + \gamma \mathbf{Z}_{dc} + \delta_{ym} + \zeta_s + \eta_z + \epsilon_{idc}
\end{equation}

\noindent
where $Mental_{idc}$ is one of the two binary indicators of severe or extreme mental health issues (depression or worry/anxiety) over the last 30 days for individual $i$ interviewed at day $d$ in PSU $c$. $Wetbulb_{dc}^{t}$ is the number of days that fell in temperature range $t$ in the previous 30 days of day $d$ in PSU $c$ ($<$16.5°C, 16.5°C-18°C, ..., $>$27°C). To be precise, the temperature bin variables will be identical for individuals that were interviewed in the same PSU at the same day, but will vary if individuals in the same PSU were interviewed on different days, or individuals were interviewed on the same day but in different PSUs. The specification further includes a series of individual and household controls $\textbf{X}$ (age, gender, years of education, marital status, rural or urban location, access to electricity, whether the individual suffers from any chronic health condition, household assets and housing characteristics). $\mathbf{Z}_{dc}$ are other weather conditions that could potentially affect mental health, specifically, I control for the rainfall deviation from the 10-year long-run rainfall in the last 30 days, the wind speed averaged over the last 30 days, and the amount of particulate matter (PM2.5) averaged over the last 30 days. Moreover, I control for the long-run (2000-2016) mean wet bulb temperature and its standard deviation averaged over the last 30 days. Thereby, the effects can be interpreted as the effect of one more heat day holding constant the temperature and its variation that would be usually observed in the given 30 day period \citep{lopalo2023temperature}. Lastly,  $\delta_{ym}$, $\zeta_s$ and $\eta_z$ are year-month, state and climate zone fixed-effects. State fixed-effects are the choice of regional fixed-effects, since health policies are commonly implemented at state/UT level in India. Year-month fixed effects account for unobserved shocks affecting mental health specific to a given year-month. Standard errors are clustered at the PSU level.  

Whereas the first empirical approach allows to determine the mental health effects at the intensive margin (i.e. one more heat day), I am also interested in the effects at the extensive margin (i.e. being exposed to a heatwave lasting multiple days). To explore these effects, I define a binary heatwave indicator $Heatwave_{dc}^{>27}$ equal to one if an individual interviewed on day  $d$ in PSU $c$ was exposed to two or more \textit{consecutive} days with a mean wet bulb temperature above 27°C. The empirical specification then reads

\begin{equation}\label{equa2}
	Mental_{idc} = \theta Heatwave_{dc}^{>27} + \vartheta \mathbf{X}_{idc} + \kappa \mathbf{Z}_{dc} + \lambda_{ym} + \mu_s + \nu_z + \varepsilon_{idc}
\end{equation}

\noindent
All control variables remain similar to equation (\ref{equa1}). I am using different temperature and daily cut-offs as well as the continuous number of consecutive heat days to test for robustness.    

To test my presumption that wet bulb temperatures have a different effect than dry bulb temperatures, I re-run all regressions with similar dry bulb indicators. Given that dry bulb temperatures are always higher than wet bulb temperatures (if humidity levels are below 100\%) and its distribution has a larger standard deviation, I employ different temperature bins and binary heatwave indicators chosen as follows. The highest temperature bin for the wet bulb specification (27°C) was chosen to correspond to the 95\textsuperscript{th} percentile of the mean wet bulb temperature distribution within the data over the three survey waves. The corresponding dry bulb temperature is 37°C, which I use to define the top bin for the dry bulb specification. As a different cut off, I choose the 99\textsuperscript{th} percentile of the dry bulb temperature distribution observed in the data, corresponding to 39°C. Additionally, to account for the larger spread of the dry bulb temperature, I use 11 instead of 9 temperature bins (with again the middle bin serving as reference category). Summary statics for the respective bins and binary heatwave indicators are presented in Table \ref{sumstats_dry} in Appendix B. They show that the chosen cut-offs generate comparable bins, with on average 1.62 days within the last 30 days falling in the highest bin with a dry bulb temperature above 37°C.

\section{Results}
\subsection{Main results}

\begin{figure}[!h]

	\includegraphics[scale=.15]{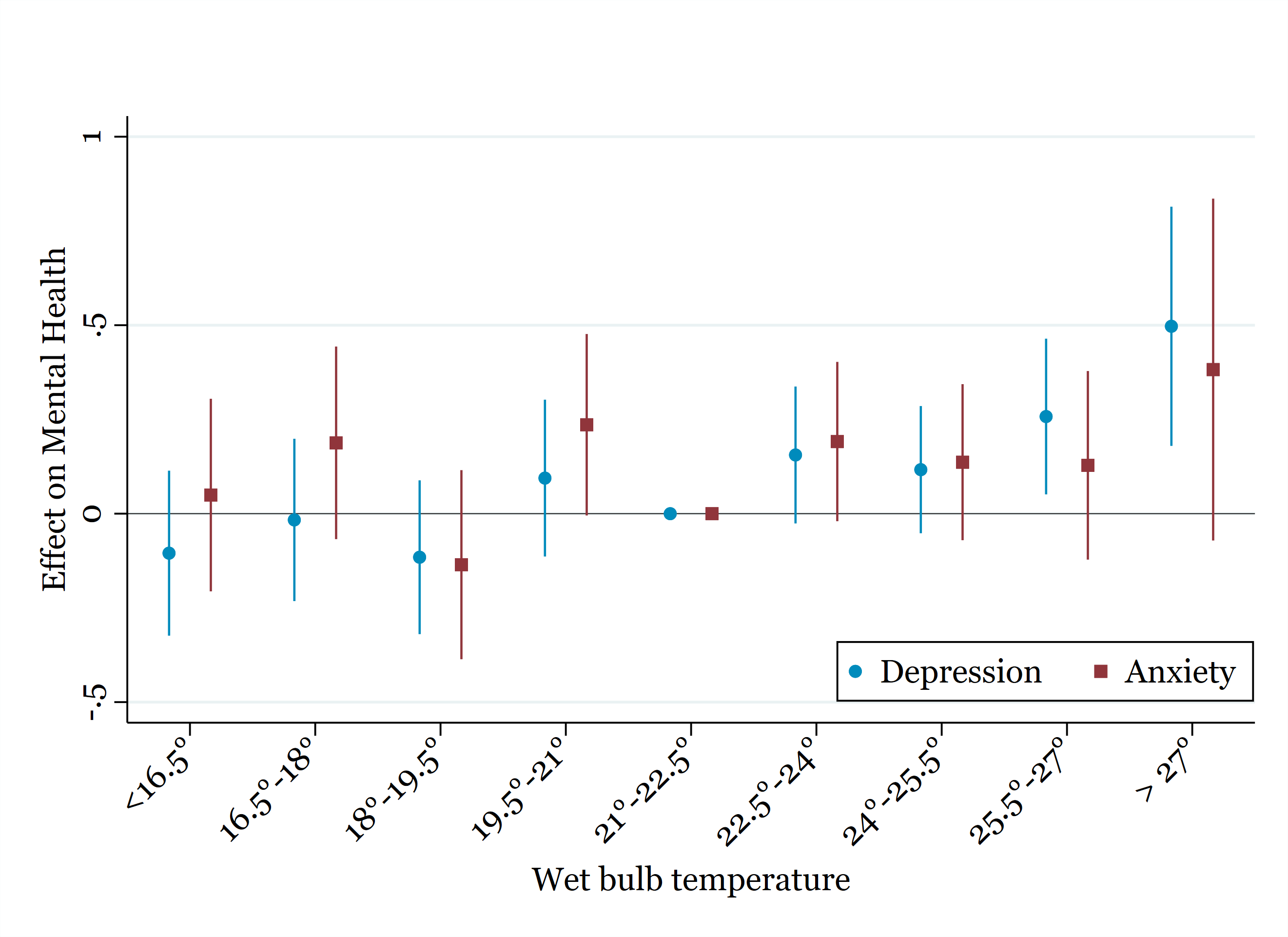}
	\centering
	\caption[Heat effects on depression and anxiety (wet bulb temperature)]{\textbf{Heat effects on depression and anxiety (wet bulb temperature)} - The figure displays the effect of the number of days in a given wet bulb temperature range relative to the number of days with a wet bulb temperature in the reference category (21°C-22.5°C) in the last 30 days on self-reported symptoms of depression and anxiety. Outcomes are binary variables (severe/extreme depression (=1); severe/extreme anxiety (=1)). Coefficients are multiplied by 100. }
	\label{figure_results}
\end{figure}

Figure \ref{figure_results} presents the results from the wet bulb temperature bin regression separately for depression (blue) and anxiety (red). Each coefficient plot is interpreted as the percentage point change in the likelihood of suffering from severe/extreme anxiety or depression that is observed if one additional day within the last 30 days had a wet bulb temperature in the range plotted on the x-axis instead of one day with a mean wet bulb temperature range between 21°-22.5°C. 

The results show that the probability of having suffered from severe/extreme depression in the last 30 days significantly increases with higher wet bulb temperatures. This is especially pronounced if wet bulb temperatures surpass 27°C: one more day above 27°C (instead of a day in the reference temperature range) increases the probability of having suffered from severe/extreme depression by 0.5 percentage points. In relative terms, this corresponds to an increase of 6\% from a mean prevalence of 8.47\%. Also for one more day in the temperature range between 25.5°C and 27°C, a significant increase of 0.25 percentage points can be observed. These results are similarly present when I use the continuous depression score (from 1-5) as outcome variable (results are displayed together with the coefficients corresponding to Figure \ref{figure_results} in Table \ref{wetbulb_R1} in Appendix C): one more day above 27°C increases the depression score by 0.017 points. The effects remain significant and become smaller (larger) when I shift the temperature bins down (up) by 1°, i.e. such that the highest bin becomes 26° (28°). The results are shown in tables \ref{wetbulb_R2} and \ref{wetbulb_R3}. 

In contrast, I do not find any significant robust effects of temperature on the probability of having suffered from extreme/severe anxiety in the last 30 days. While the point estimates for two top bins are positive, they are not statistically significant and also some of the lower bins point in a positive direction, rejecting a clear relationship between high wet bulb temperatures and anxiety levels. There is also no significant effect when the continuous anxiety score is used as outcome variable (Table \ref{wetbulb_R1}). For this reason, I focus in the remainder solely on depression.

Table \ref{results} presents the results for depression using alternative heat indicators.\footnote{The results for anxiety using alternative heat indicators are shown in Table \ref{results_anxiety} in Appendix C.} Column (1) uses only the number of days above 27°C as explanatory variable (without any other temperature bins) and confirms the main results. One more heat day above 27° in the last 30 days increases the likelihood of having suffered from extreme or severe depression by 0.3 percentage points. In Column (2), I focus on the number of \textit{consecutive} days above 27°C. The results show that consecutive heat days indeed are worse for mental health than the same number of heat days which are not necessarily consecutive;  if the number of consecutive days above 27° increases by one day, it increases the likelihood of having suffered from extreme or severe depression by 0.64 percentage points. Column (3) presents the main results for the binary heatwave indicator; individuals that were exposed to two or more consecutive days with a mean wet bulb temperature above 27°C were two percentage points more likely to have suffered from severe/extreme depression in the last 30 days. If instead three or more days are used as the cut-off point, this effect increases to 2.3 percentage points. In relative terms, these numbers correspond to an increase of 24\% and 27\% relative to the mean prevalence of severe/extreme depression. When a heatwave is defined as 2 or more consecutive days above 28°C (corresponding to the 99\textsuperscript{th} percentile of the wet bulb temperature distribution in the data sample), the effect becomes as large as 5.8 percentage points. However, such extreme heat events are very rare. Only 2.1\% of the individuals in the sample experienced such a heatwave in the last 30 days. 

\pagebreak

\begin{table}[]
		\caption{\label{results}Alternative heat indicators - Wet bulb}
		\small
		\centering
	\begin{tabular}{lccccc}
		\midrule
		\midrule
				& (1)	& (2)	& (3) 	& (4)	& (5) \\
		& Sev./Extr.	& Sev./Extr. 	& Sev./Extr. 	& Sev./Extr.	& Sev./Extr. \\ 
		& Depression 	& Depression 	& Depression 	& Depression 	& Depression \\
		\midrule
\# Days $\geq$ 27°C (w/o other bins)  & 0.316*** &         &         &         &          \\
& (0.118)  &         &         &         &          \\
\# Consecutive days $\geq$27°C    &          & 0.642*** &         &         &          \\
&          & (0.209)  &         &         &          \\
Heatwave ($\geq$2 days $\geq$27°C) &          &         & 1.994** &         &          \\
&          &         & (0.931) &         &          \\
Heatwave ($\geq$3 days $\geq$27°C) &          &         &         & 2.324** &          \\
&          &         &         & (1.024) &          \\
Heatwave ($\geq$2 days $\geq$28°C)         &          &         &         &         & 5.880** \\
&          &         &         &         & (2.538)  \\
&          &         &         &         &          \\
Observations         & 27,287   & 27,287  & 27,287  & 27,287  & 27,287  \\
R-squared            & 0.070    & 0.070   & 0.069   & 0.069   & 0.070    \\
Weather Controls     & \Checkmark      & \Checkmark       & \Checkmark       & \Checkmark       & \Checkmark        \\
Ind. Controls     & \Checkmark      & \Checkmark       & \Checkmark       & \Checkmark       & \Checkmark        \\
Year-month FE     & \Checkmark        & \Checkmark       & \Checkmark       & \Checkmark       & \Checkmark        \\
State FE     & \Checkmark        & \Checkmark       & \Checkmark       & \Checkmark       & \Checkmark        \\
KGC FE               & \Checkmark        & \Checkmark       & \Checkmark       & \Checkmark       & \Checkmark        \\
\# of PSU            & 371      & 371     & 371     & 371     & 371     \\
Mean of dependent variable            & 8.47\%      & 8.47\%      & 8.47\%      & 8.47\%      & 8.47\%   \\
		\midrule
		\midrule
		\multicolumn{6}{p{15.5cm}}{\textit{Notes:} Robust standard  errors are clustered at the PSU level and presented in parentheses. ***   p\textless{}0.01, ** p\textless{}0.05, * p\textless{}0.1.}
	\end{tabular}
\end{table}

\subsection{Dry bulb temperature}
\begin{figure}[!h]
	\includegraphics[scale=.25,trim={0 6cm 0 0},clip]{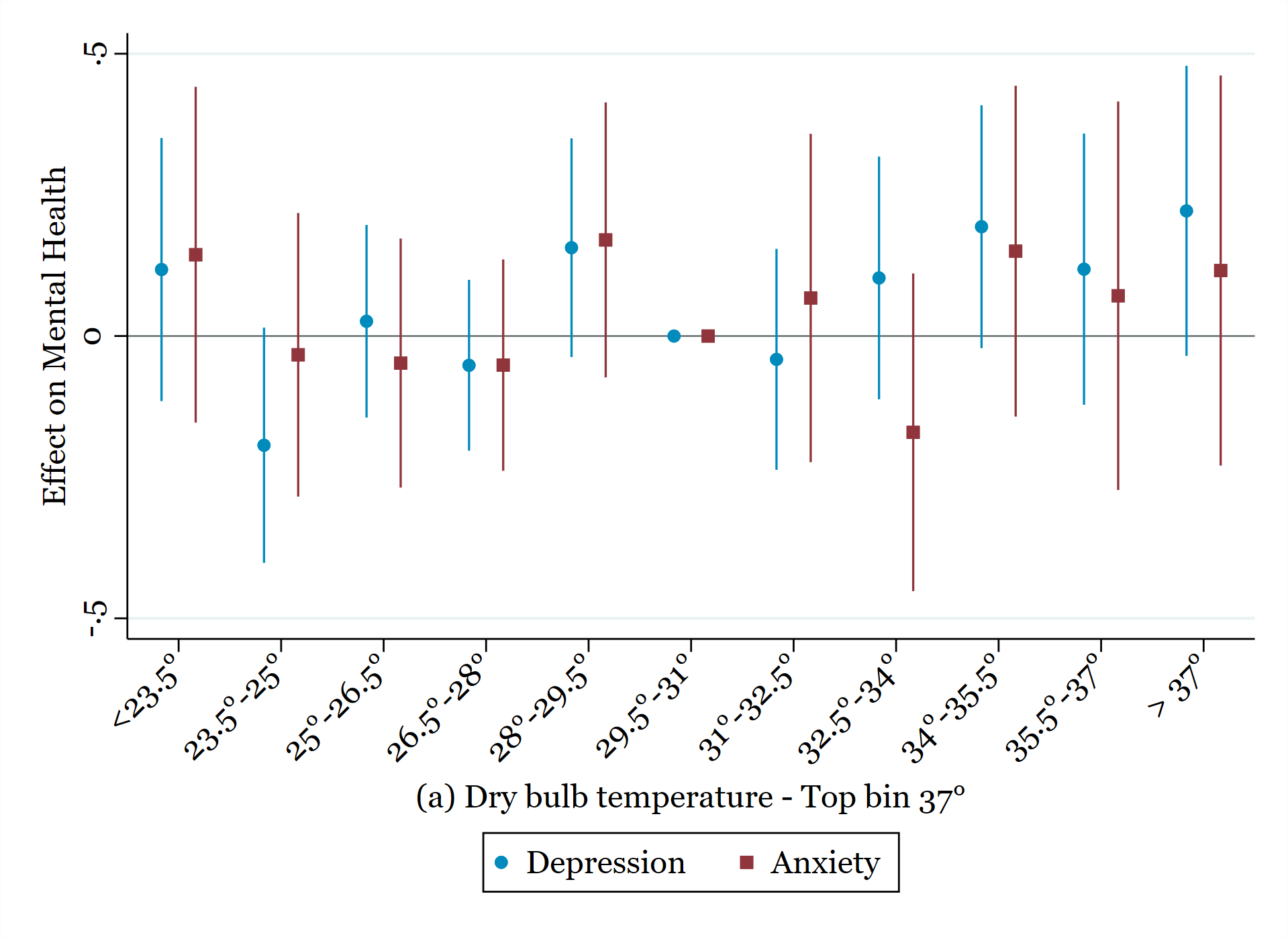}
	\includegraphics[scale=.25]{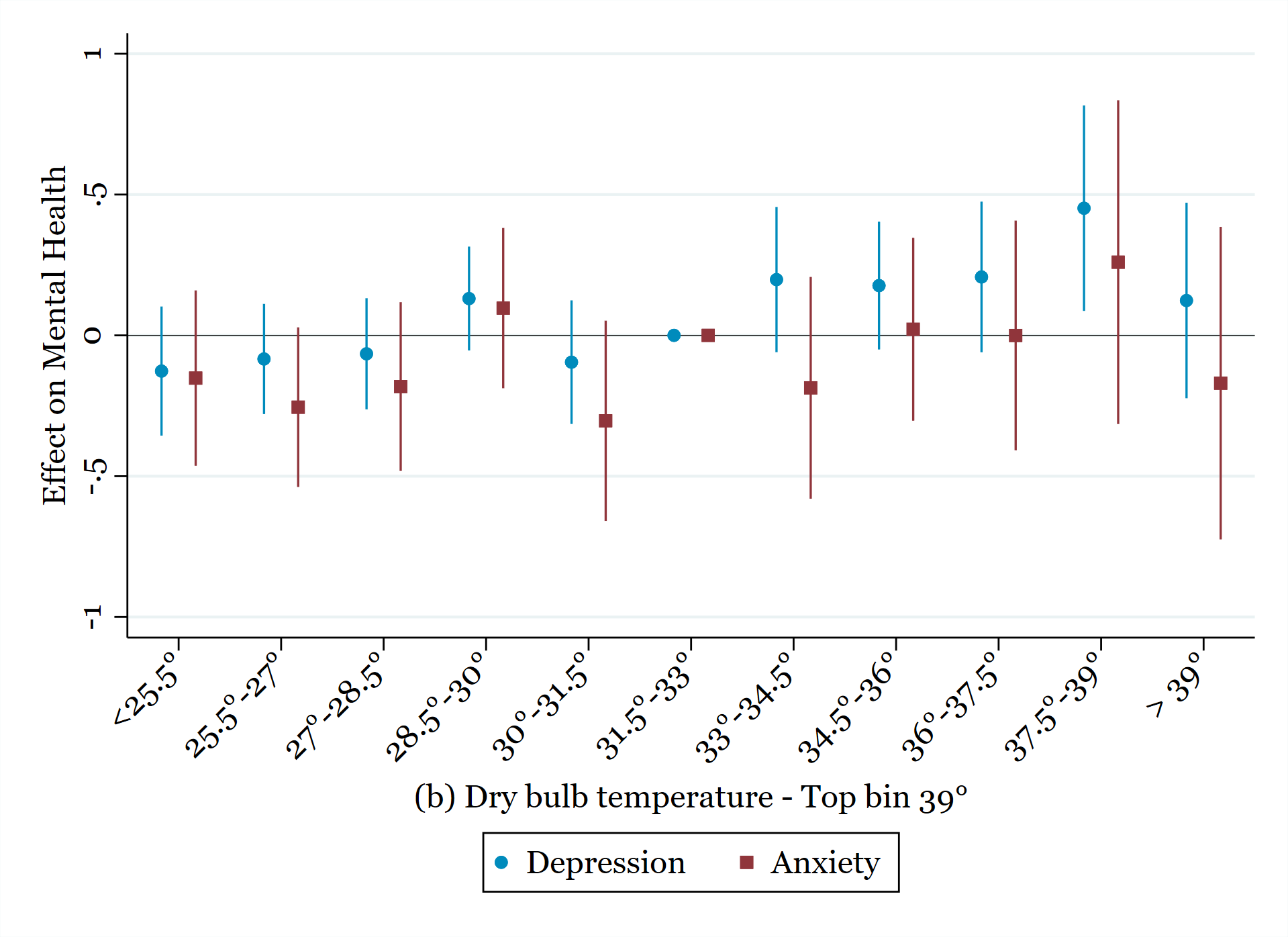}
	\centering
	\caption[Heat effects on depression and anxiety (dry bulb temperature)]{\textbf{Heat effects on depression and anxiety (dry bulb temperature)} - The figure displays the effect of the number of days in a given dry bulb temperature range relative to the number of days with a dry bulb temperature in the reference category (29.5°C-31°C /31.5°C-33°C) in the last 30 days on self-reported symptoms of depression and anxiety. Outcomes are binary variables (severe/extreme depression (=1); severe/extreme anxiety (=1)). Coefficients are multiplied by 100. }
	\label{figure_results_temp}
\end{figure}

Having established that exposure to high wet bulb temperatures increases the likelihood of suffering from severe/extreme depression, I now turn to the effects of dry bulb temperatures. Figure \ref{figure_results_temp} displays the results for the temperature bin regressions with dry bulb temperature as main explanatory variable. In graph (a), the highest dry bulb temperature bin corresponds to 37°C and in graph (b), the highest dry bulb temperature bin corresponds to 39°C. In both regression, average relative humidity over the last 30 days is included only as ordinary control variable.

There is no  strong relationship between the dry bulb temperature bins and both mental health outcomes. For depression, the higher bins point towards an increase in the prevalence of severe/extreme depression, but the highest bins are not not or only marginally significant. Only the coefficient for the temperature bin 37.5C°-39°C reaches significance at conventional levels. Yet, the effect size for this bin is half as large as the effect of the highest bin in the wet bulb specification. 

Similarly, almost all alternative dry bulb heat indicators show only a marginally significant or insignificant relationship with the depression outcome, as shown in Table \ref{results_dry}. Only when using the binary heatwave indicator of 3 or more consecutive days above 37°C, I detect an increase of 2.5 percentage points in the prevalence of severe/extreme depression -- half as large as the effect found for the wet bulb heatwave indicator. 

This suggests that earlier studies might have indeed underestimated the relationship between heat exposure and mental health. This could also explain some of the null-results of dry bulb heat on depression as found in \cite{pailler2018effects}.

\begin{landscape}
	
	\begin{table}[]
		\caption{\label{results_dry}Alternative heat indicators - Dry  bulb}
		\small
		\centering
		\begin{tabular}{lcccccccc}
			\midrule
			\midrule
			& (1)	& (2)	& (3)	& (4)	& (5) & (6)	& (7)	& (8) \\ 
			& Sev./Extr.	& Sev./Extr. 	& Sev./Extr. 	& Sev./Extr.	& Sev./Extr. & Sev./Extr.	& Sev./Extr. 	& Sev./Extr. \\ 
			& Depression 	& Depression 	& Depression 	& Depression 	& Depression 	& Depression 	& Depression & Depression \\
			\midrule
			\# Days $\geq$ 37°C (w/o other bins) & 0.097 &         &         &         &       &   &   & \\
			& (0.078) &         &         &         &          &   &  &  \\
			\# Days $\geq$ 39°C (w/o other bins)  &          & -0.013 &         &         &           &   &  &  \\
			&          & (0.137) &         &         &           &   &   & \\
			Heatwave ($\geq$2 days $\geq$37°C) &          &         & 1.654* &         &          &   &  &  \\
			&          &         & (0.994) &         &          &   &  &  \\
			Heatwave ($\geq$3 days $\geq$37°C) &          &         &         & 2.536** &          &   &   & \\
			&          &         &         & (1.007) &           &   & &   \\
			Heatwave ($\geq$2 days $\geq$39°C)         &          &         &         &         & 1.289   &   &  &  \\
			&          &         &         &         & (1.364)    &   &   & \\
			Heatwave ($\geq$3 days $\geq$39°C)	&          &         &         &         &           &  -0.495 &  &  \\
			&          &         &         &         &           & (1.215)  &  &  \\
			\# Consecutive days $\geq$37°C		&          &         &         &         &           &   & 0.194* &  \\
			&          &         &         &         &           &   &  (0.107) &  \\
			\# Consecutive days $\geq$39°C		&          &         &         &         &           &   &  & -0.003 \\
			&          &         &         &         &           &   &  & (0.223) \\

			Observations         &27,287   & 27,287 & 27,287  & 27,287  & 27,287    & 27,287   & 27,287   &  27,287\\
			R-squared            & 0.069    & 0.069   & 0.070   & 0.070   & 0.069      &  0.069    &  0.070   &  0.070  \\
			Weather Controls     & \Checkmark      & \Checkmark       & \Checkmark       & \Checkmark       & \Checkmark      & \Checkmark       & \Checkmark &   \Checkmark    \\
			Ind. Controls     & \Checkmark      & \Checkmark       & \Checkmark       & \Checkmark       & \Checkmark      & \Checkmark       & \Checkmark &  \Checkmark     \\
			Year-month FE     & \Checkmark        & \Checkmark       & \Checkmark       & \Checkmark       & \Checkmark        & \Checkmark       & \Checkmark  & \Checkmark     \\
			State FE     & \Checkmark        & \Checkmark       & \Checkmark       & \Checkmark       & \Checkmark        & \Checkmark       & \Checkmark  & \Checkmark     \\
			KGC FE               & \Checkmark        & \Checkmark       & \Checkmark       & \Checkmark       & \Checkmark       & \Checkmark       & \Checkmark &   \Checkmark    \\
			\# of PSU            & 371      & 371     & 371     & 371     & 371   & 371 &371 &371  \\
			Mean of dependent variable               & 8.47\%      & 8.47\%      & 8.47\%      & 8.47\%      & 8.47\%  & 8.47\% & 8.47\% & 8.47\%  \\

			\midrule
			\midrule
			\multicolumn{8}{p{20cm}}{\textit{Notes:} Robust standard  errors are clustered at the PSU level and presented in parentheses. ***   p\textless{}0.01, ** p\textless{}0.05, * p\textless{}0.1.}
		\end{tabular}
	\end{table}

\end{landscape}

\subsection{Robustness checks and dynamics}

I conduct a series of robustness checks for both wet bulb temperature specifications; the temperature bin specification and the specification using the binary heatwave indicator. The results are presented in Table \ref{robustness} in Appendix D. For readability, I only show the coefficient for the highest temperature bin (>27°) for the bin regression, but in all regressions, the full set of bins is included. 

Column (1) shows once again the baseline specification. In Column (2), district fixed-effects (administrative level 2) instead of state fixed-effects (administrative level 1) are included in the regression to verify that the results are not driven by heterogeneity across districts. In Column (3), I include year and month fixed-effects separately instead of year-month fixed-effects, allowing for month-specific reoccurring unobserved shocks that could affect depression outcomes. In Column (4) and (5) I drop the state and climate zone fixed-effects, respectively, to verify that results are not driven by the choice of fixed-effects and whether regional heterogeneity might bias the results. In Column (4), I drop all other weather variables and in Column (5) I drop all individual and household controls to verify that the effects are not dependent on the choice of covariates, and lastly in Column (6), I use a logit model specification instead of a linear probability model to account for the bounded nature of the binary outcome variable. 

Overall, the results are very robust to these changes in the specifications, though the coefficient sizes slightly change in size in some specifications. Specifically, effect sizes become particularly large when no state or district fixed-effects are included (Column(4)), suggesting that there are indeed unobserved geographical factors that correlate with mental health outcomes. In contrast, effect sizes are smallest when other weather and climate conditions (rainfall, wind speed etc.) are not controlled for (Column (6)), highlighting the importance of the interplay between different climate conditions. 

I next turn to the temporal dynamics of the heat effect on mental health. Specifically, I am interested in whether the effect of high wet bulb temperatures on depression is solely contemporaneous, or also measurable over a longer time horizon. To explore these dynamics, I change the horizon over which I count the number of days within the different temperature bins. The results are shown in Table \ref{wetbulb_dynamics}. In Column (1), the bins are constructed only over the last two weeks before the respective interview took place ($day_{t-1}-day_{t-14}$), in Column (2), the horizon is shifted to week three and four before the interview ($day_{t-15}-day_{t-30}$), Column (3) focuses on the complete second to last month ($day_{t-31}-day_{t-60}$), and lastly, Column (4) serves as placebo test and uses the temperature distribution of the following month to construct the temperature bins. Overall, the results suggest that the effect is not limited to a short time horizon, but sustained over a longer period of time. Specifically, the coefficients for the highest temperature bin remain large and significant not only when focusing on the last 14 days before the interview, but even high temperatures two weeks and more than one months ago have a noticeable effect on the current mental health state. Lastly, the placebo specification in Column (4) shows no statistically significant effect, confirming that the actual effects are not driven by statistical chance.

\newpage
\subsection{Heterogeneity}
\begin{figure}[!h]
	\includegraphics[scale=.4]{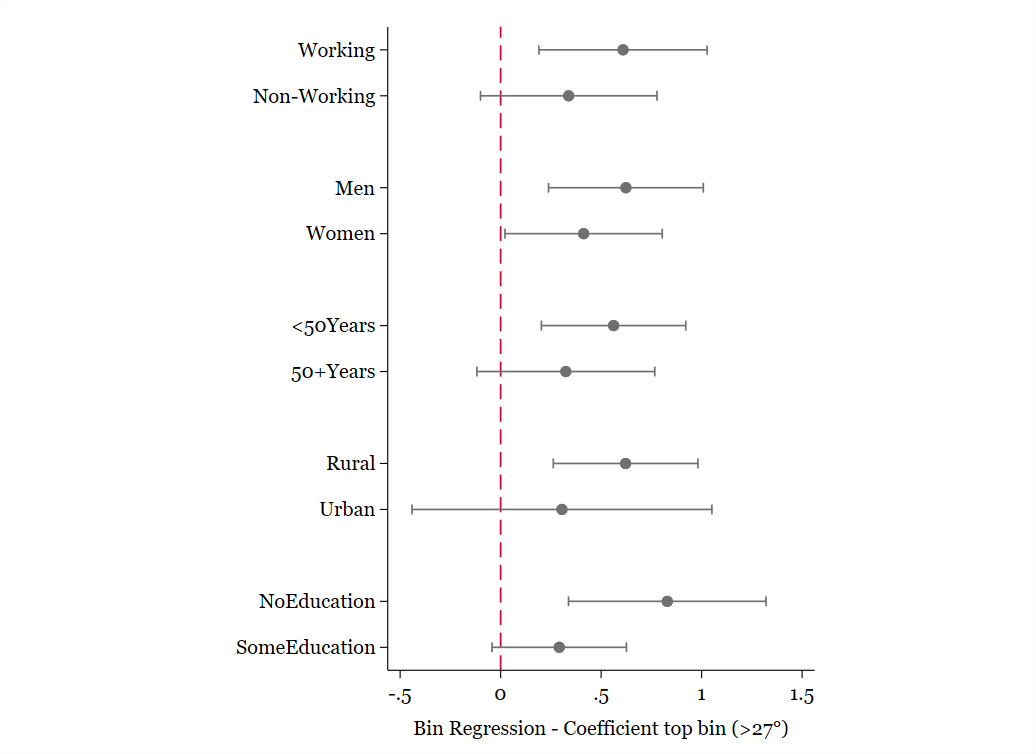}
	\centering
	\caption[Heterogeneous effects]{\textbf{Heterogeneous effects} - The figure displays the heat effect for different socioeconomic subgroups using the temperature bin regression. Coefficients are from the top bin (>27°C). The outcome variable is binary and multiplied by 100, hence, coefficients are directly interpretable as percentage points.}
	\label{hetero}
\end{figure}

In the following, I assess the effects of heat exposure on mental health separately by working status, gender, age, place of living and education level to identify vulnerable populations. Figure \ref{hetero} presents the heterogeneous effects by these socioeconomic characteristics using the temperature bin regression approach. The coefficients represent the effect for the highest bin (>27°), stemming from the temperature bin regression.

The analyses are well in line with previous studies indicating that working individuals, rural populations and those with lower education status are more prone to heat effects \citep{carleton2017crop, pailler2018effects, liu2019influence, hua2023effects}. In particular, in the context of India or other LMICs, individuals with lower education and in rural areas are more likely to engage in agricultural activities or other physically demanding jobs outside, often have less resources to engage in adaptive measures and are hence more exposed to extreme temperatures. Figure \ref{hetero} shows that the point estimates for working individuals, those living in rural areas and those with no education are larger than their comparison counterparts. However, only in the case of education, the point estimates are also statistically different from each other, calling for caution to interpret this as robust evidence for differences with respect to working status and place of living. In particular, in terms of place of living, the sample for the urban population is small in comparison to the rural sample (N=6,751 vs. N=20,535), which could partly explain the large confidence intervals. 

In terms of heterogeneous heat effects on mental health by gender, the existing literature is inconclusive. Whereas studies investigating heat effects on suicide rates mostly find larger effects for men \citep{thompson2018associations, liu2021there}, studies on self-reported mental health outcomes often find larger effects for women \citep{hua2023effects, obradovich2018empirical}. My results suggest that men and women are not differently affected by heat. While the point estimates are slightly larger for men, the effects are not statistically different from each other, which would allow for a definite conclusion. 

However, my results are in strong contrast to other studies when it comes to the heterogeneous effects by age. While previous studies find that especially elderly are prone to heat, also in terms of mental health outcomes \citep{hua2023effects, zhang2023heatwave}, my results show that the effect is larger for younger cohorts below the age of 50, and the effect for those above the age of 50 is not statistically significant. A potential reason could be that younger individuals are more likely to work and the age effect hence only reflects the working effect. However, the age difference remains even when limiting the sample to younger and older working individuals; the heat effect is larger and significant for young workers but not significant for older workers. Moreover, also for younger non-working individuals, the effect remains similar in size and significance. Also, power issues cannot explain the differential effect since samples sizes for those below and above the age of 50 are almost equal (with the elderly sample being even larger). Another reason could be that the elderly sample contains many individuals that generally live longer than the average of their respective cohorts; the mean age in the 50+ sample is 62 and 36\% of the individuals are 65 or older. Under the presumption that these individuals are generally healthier than the average (given that for cohorts born in India in the 1940s and 1950s, life expectancy at birth was about 40 years and only about 35\% of these cohorts reached the age of 65 \citep{worldbank2025, UNdata2025}), they also might be less prone to suffer under heat. And alternative explanation could be that mental health stigma is lower among younger cohorts, and hence younger individuals are more likely to truthfully report mental health outcomes \citep{bharadwaj2017mental}. However, what speaks against this hypothesis is that the mean prevalence of depression among those individuals below the age 50 in the sample is on average lower (7.3\%) than the mean prevalence for those above the age of 50 (9.4\%). Hence, while the reason remains purely speculative, the findings are reason to concern; if especially younger and productive individuals are those most likely to suffer mentally from increasingly more frequent heatwaves, this can lead to large productivity losses in the future \citep{bubonya2017mental}.

\newpage
\section{Channels}
I next investigate different channels through which heat exposure can affect mental health. First, high temperatures during the night can worsen sleeping patterns and thereby cause or exacerbate depressive symptoms and worsen mental well-being \citep{obradovich2017nighttime, mullins2019temperature, carias2024temperature}. Second, heat can also increase aggression, conflict and violence, which in turn are harmful to mental health \citep{hsiang2013quantifying, baylis2020temperature, mukherjee2021causal}. In the following, I explore in turn whether these two channels are of relevance in the given context.

Within the scope of WHO-SAGE information about the level of conflict and tension with others over the last 30 days as well as about sleeping problems over the last 30 days were collected, allowing to assess whether these two factors function as channels of the heat-mental health relationship. The questions for conflict and tension  as well as for sleeping problems were designed similar to the depression and anxiety questions (``Overall in the last 30 days, how much difficulty did you have in dealing with conflicts and tensions with others? / Overall in the last 30 days, how much of a problem did you have with sleeping, such as falling asleep, waking up frequently during the night or waking up too early in the morning?''). Hence, similar to the main outcomes, I construct binary indicators equal to one if the individual reported to had encountered severe/extreme conflicts and tensions or severe/extreme sleeping problems in the last 30 days and regress the indicators on the wet bulb heat indicators. The results are presented in Table \ref{channels} and show that conflict and tensions tend to increase with extreme temperatures, whereas I find no effects on reported sleeping difficulties. This suggest that increased conflict and tension might be one channel through which extreme temperatures drive deterioration of mental health. This is further consolidated when including the conflict variable as moderating variable in the main depression regression; the effect size of heat on depressions is reduced by about 15\% (both for the bin regression and binary heatwave indicator), while the correlation between conflict and depression is statistically significant (not shown in Table). 

Heatwaves during the main agricultural season are another likely channel of the heat-mental health nexus, since physically demanding activities, such as sowing or harvesting, are conducted during extreme heat \citep{berry2010climate, el2022impacts}. Moreover, extreme heat during the main agricultural season might be correlated with the occurrence of droughts, which in turn might increase expectations of an adverse harvest outcome and poverty \citep{carleton2017crop}. Hence, both physiological mechanisms  as well as economic considerations can be a driving force of the heat-mental health nexus. While disentangling the precise mechanism remains difficult, the focus of my analysis on exposure to heat events in the last 30 days renders a physiological mechanism the more plausible explanation than economic expectations which might only unfold in the longer-run. Moreover, since crops that are planted during the main agricultural season (Kharif) are more sensitive to water shortages than high temperatures, and rainfall levels are held constant in the analysis, the effect captured by the wet bulb temperature rather reflects the physiological channel of exposure to heat stress on mental health. 

In Table \ref{channels2}, I split the sample by urban and rural population and interact the temperature bins and the binary heatwave indicator with a binary variable for the main Indian agricultural season Kharif. Kharif lasts from June to October and crops are sown when the Indian south-west monsoon rains start. The results in Column (1) and (2) show that the non-significant effect of wet bulb temperature on mental health in urban areas is not heterogeneous over the agricultural cycle. In contrast, in rural areas, the mental health consequences due to high wet bulb temperatures are stronger in the Kharif season. In fact, the linear coefficients for both heat indicators even turn negative (implying an improvement in mental health), whereas the interaction term is large in absolute terms and statistically significant. Hence, the effect of  high wet bulb temperatures on the likelihood of suffering from severe or extreme depression is largest during the main agricultural season, suggesting that continued exposure to heat due to outside and physically demanding activities amplifies the effect of high wet bulb temperatures on mental health.

\begin{table}[h!]
	\caption{\label{channels}Channels - Conflict and sleep}
	\centering
	\small
\begin{tabular}{lC{1.5cm}C{1.5cm}C{1.5cm}C{1.5cm}C{1.5cm}C{1.5cm}}
		\midrule
	\midrule
  & (1)     & (2)      & (3)        & (4)     & (5)      & (6)        \\
& Conflict   & Conflict & Conflict & Sleep   & Sleep & Sleep \\
\midrule
&         &          &            &         &          &            \\
\# Days \textgreater 27°C   & 0.298*  &         &          & 0.152   &         &         \\
& (0.156) &         &          & (0.147) &         &         \\
Heatwave ($\geq$2 days $\geq$27°)   &         & 1.374   &          &         & -0.597  &         \\
&         & (1.255) &          &         & (0.982) &         \\
Heatwave ($\geq$2 days $\geq$28°)  &         &         & 5.724*** &         &         & 1.517   \\
&         &         & (1.983)  &         &         & (1.870) \\
& & & & & & \\
Observations         & 27,278  & 27,278  & 27,278   & 27,300  & 27,300  & 27,300  \\
R-squared            & 0.079   & 0.078   & 0.079    & 0.074   & 0.074   & 0.074   \\
\# of PSU            & 371      & 371     & 371     & 371     & 371   & 371  \\
Mean of dependent variable               & 9.66\%      & 9.66\%         & 9.66\%         & 9.22\%         & 9.22\%     & 9.22\%    \\
	\midrule
	\midrule
	\multicolumn{7}{p{16cm}}{\textit{Notes:} All regressions include weather controls, individual controls, year-month fixed-effects, state  fixed-effects and climate zone fixed effects. Regressions for temperature bins (Columns (1) and (4)) include all temperature bins but only the top bin is shown in the table. Robust standard errors are clustered at the PSU level and shown in parentheses. *** p\textless{}0.01, ** p\textless{}0.05, * p\textless{}0.1.}
	\end{tabular}
\end{table}

\begin{table}[!]
		\caption{\label{channels2}Channels - Kharif season}
	\centering
	\small
	\begin{tabular}{lcccc}
		\midrule
		\midrule
		          &       (1)  &    (2)    &   (3)       &  (4)       \\
				& Sev./Extr.	& Sev./Extr. 	& Sev./Extr. 	& Sev./Extr.	 \\ 
	& Depression 	& Depression 	& Depression 	& Depression  \\
		         
		         &    Urban     &      Urban   &     Rural     &     Rural     \\ \hline
                                 &         &         &          &          \\
\# Days \textgreater 27°C                & -1.134  &         & -1.272** &          \\
& (1.056) &         & (0.511)  &          \\
\# Days \textgreater 27°C  * Kharif     & 1.538   &         & 1.992*** &          \\
& (0.992) &         & (0.482)  &          \\
Heatwave ($\geq$2 days $\geq$27°)             &         & -2.331  &          & -4.621** \\
&         & (3.232) &          & (2.017)  \\
Heatwave ($\geq$2 days $\geq$27°) * Kharif &         & 2.969   &          & 8.083*** \\
&         & (3.359) &          & (2.045)  \\
&         &         &          &          \\
Observations                     & 6,751   & 6,751   & 20,535   & 20,535   \\
R-squared                        & 0.107   & 0.104   & 0.067    & 0.064    \\
\# of PSUs  &  89       &      89   &      282    &      282    \\
Mean of dependent variable    & 6.76\% & 6.76\%  & 9.03\% & 9.03\%\\
		\midrule
		\midrule
		\multicolumn{5}{p{14cm}}{\textit{Notes:} All regressions include weather controls, individual controls, year-month fixed-effects, state fixed-effects and climate zone fixed effects. The temperature bin regressions (Columns (1) and (3)) include all bins and its interactions with the Kharif dummy, but on the top bin is shown in the table. The Kharif dummy is absorbed by the year-month fixed effects. Robust standard errors are clustered at the PSU level and presented in parentheses. *** p\textless{}0.01, ** p\textless{}0.05, * p\textless{}0.1.}
	\end{tabular}
\end{table}

\newpage

\section{Effectiveness of the Indian District Mental Health Program in reducing negative heatwave effects}
This section explores whether an existing mental health program in India has the potential to effectively counteract the negative mental health impacts of heatwaves. Since the early 1980s, the Indian Ministry of Health and Family Welfare started to implement programs and initiatives to promote mental health among its population. The National Mental Health Program (NMHP) was initiated in 1982 and as part of this program the District Mental Health Program (DMHP) was integrated into the national program in 1996 \citep{girase2022india, gangadhar2023mental}. The DMHP targeted the implementation of community-based mental health services, including the training of community mental health workers, early detection and treatment programs as well as awareness generation and the integration of mental health interventions and programs within the public health system \citep{dghs}.

Starting as a pilot project in four districts in 1996, it has since then been rolled out in a staggered design across almost the entire country and is predicted to cover 767 districts by the end of 2025 \citep{mohfw, gangadhar2023mental}. By the end of 2014, 32 districts within the WHO data sample were covered under the DMHP; 4 already before Wave 0, 16 were covered in between Wave 0 and Wave 1, and 12 in between Wave 1 and Wave 2 (Table \ref{DMHP}, Appendix E, extracted from \cite{dghs}). In the following, I assess whether the program was able to counteract some of the previously recorded negative mental health effects of heatwaves by providing improved access to mental health care and increased awareness about mental health issues. To do so, I modify equation (1) and (2) to include a binary indicator equal to one if district \textit{i} was covered under the DMHP in year \textit{t}, and its interaction term with the temperature bins or heatwave indicator, respectively. Moreover, I include district-fixed effects in the specification to account for the program being rolled out on the district- and not state-level. The linear term of the DMHP indicator then represents the general (difference-in-difference) correlation between DMHP availability and mental health outcomes, while the interaction term can be interpreted as the mediating effect of program availability on the heatwave-mental health link. Note that this should not be interpreted as causal effect, but rather as correlational evidence, given that the roll-out of the program might have been targeted to purposely chosen districts first.

The results of this analyses are presented in Table \ref{MHP} -- with the bin specification in Column (1) and the binary indicator in Column (2). The coefficient for the linear DMHP term is large in absolute terms and significant, suggesting that the program indeed was rolled out first to districts where rates of depression (or generally rates of mental illness) were high to begin with. Also, the heat indicators remain significant, with effect sizes very similar to those in the main results. The main coefficients of interest are the two interaction terms between the heat indicators and the DMHP variable. Both have a negative sign and the coefficient in the bin regression reaches statistical significance. Hence, it appears that the mental health consequences of extreme heat are to some extent alleviated if individuals have access to the DMHP, suggesting a protective role of the program in mitigating adverse mental health effect. This also speaks in favor of integrating specifically climate-sensitive mental health components within the existing framework of the DMHP. Nevertheless, given the steady increase in the coverage rate of the program, it remains unanswered whether this alleviating effect remained once the program was rolled-out to the entire country, which the current data do not allow to address.

\begin{table}[h!]
	\caption{\label{MHP} District Mental Health Program}
	\centering
	\small
	\begin{tabular}{lcc}
		\midrule
		\midrule
		& (1)                   & (2)                 \\
        & Sev./Extr.  & Sev./Extr. . \\
	     & Depression &Depression  \\
	     \midrule
DMHP                      & 5.373* &	3.050*** 
    \\
& (3.078) &	(1.006)    \\
\# Days \textgreater 27°C              & 0.363**  &                   \\
& (0.184)   &                    \\
\# Days \textgreater 27°C  * DMHP     & -0.558** &                 \\
& (0.238)   &                    \\
Heatwave ($\geq$2 days $\geq$27°)         &           & 2.182*
             \\
&            &      (1.216)     \\
Heatwave ($\geq$2 days $\geq$27°) * DMHP &           & -0.362
          \\
&           &   (1.905)        \\
& & \\

Observations                  & 27,287    & 27,287      \\
R-squared                     & 0.087     & 0.085      \\
District fixed-effects & \Checkmark & \Checkmark \\
\# of PSU                     & 371       & 371         \\
Mean of dependent variable               & 8.47\%      & 8.47\%     \\
		\midrule
		\midrule
	\multicolumn{3}{p{12.5cm}}{\textit{Notes:} All regressions include weather controls, individual controls, year-month fixed-effects, district fixed-effects and climate zone fixed-effects. The temperature bin regression (Column (1)) includes all temperature bins and their interactions with the DMHP indicator, but only the top bin is shown in the table. Robust standard errors are clustered at the PSU level and shown in parentheses. *** p\textless{}0.01, ** p\textless{}0.05, * p\textless{}0.1.}
	\end{tabular}
\end{table}

\pagebreak
\section{Discussion and conclusion}
\noindent
This study investigates the consequences of extreme heat and humidity on self-reported mental health outcomes in India. Yet, in contrast to previous studies, it specifically considers the non-linear interaction between temperature and humidity, as captured by the wet bulb temperature, which provides a more accurate measure of heat stress than measures relying on the conventional dry bulb temperature. My results reveal that exposure to wet bulb heat significantly increases the likelihood of severe depression. Specifically, using multiple heat indicators to capture effects at both the extensive and intensive margin, I find that a heatwave in the 30 days preceding the survey increases the probability of suffering from severe or extreme depression by 24\% at the extensive margin (exposure to a heatwave lasting two or more days) and by 6\% at the intensive margin (one additional heat day). 

The adverse effects on depression are pronounced when considering the interaction of temperature and humidity, whereas the effects are considerably smaller in size and mostly non-significant when relying on the conventional dry bulb temperature. This suggests that  studies focusing solely on temperature may have underestimated the true impact of heat on mental health, calling for a broader inclusion of humidity in climate-health research frameworks. Furthermore, my analysis highlights that vulnerable groups -- in particular rural populations, those employed in agricultural activities, or those with lower educational attainment -- are disproportionately affected. This is in line with previous findings for the heat-health link in India \citep{banerjee2020heat, pailler2018effects} and underscores the necessity of adopting targeted interventions to avoid that existing inequalities are exacerbated by climate change.

Moreover, given the observation of the temporal persistence of the heat effects on mental health, heatwaves do not lead to merely acute deterioration in mental health but may have longer-lasting implications. This implies that potential policies aiming to address mental health consequences should not only be short-term focused. It also highlights the need for sustained mental health monitoring and adaptive strategies that extend beyond immediate responses to heatwaves. Lastly, in contrast to findings from previous studies \citep[e.g.,][]{hua2023effects, zhang2023heatwave}, my heterogeneity analysis reveals that the mental health impacts of heat are amplified among younger cohorts. This can lead to potential long-term productivity losses and also highlights the need for adaptive policies from an economic perspective. 

In contrast to previous research \citep{obradovich2018empirical, hansen2008effect}, I do not find any significant effects on anxiety levels. However, the absence of significant effects in this study may stem from the nature of anxiety as a mental health condition. Anxiety often manifests as an anticipatory response, rooted in fear or concern about future events \citep{apa2025}. Unlike depression, which may be more directly tied to immediate mood disruptions and physiological stress caused by heat, anxiety might not emerge strongly in response to short-term heat exposure. This distinction aligns with the existing literature, which highlights increases in anxiety linked to long-term climate change rather than acute heat stress; the concept of ``climate anxiety'', for example, captures the psychological burden of grim future climate scenarios and the perceived helplessness in addressing them \citep{wu2020climate, clayton2020development}. Hence, while depression and anxiety are both mental health outcomes influenced by environmental stressors, their triggers and manifestations differ, with anxiety potentially requiring a longer time horizon or more indirect pathways to become salient.

While the study establishes a robust association between wet bulb temperature and depression, limitations remain in terms of explicitly disentangling the precise pathways. While I touch upon the roles of social conflict and exposure to heat during harvesting seasons, future research could aim for exploring the links between heat and conflict more closely. Similarly, disentangling whether heat shocks during the main agricultural season deteriorate mental health mostly by physiological mechanisms or by economic consequences and considerations, could advance the existing literature.     

The significant relationship between heat and depression calls for the expansion of mental health services under the looming consequences of climate change. The integration of climate-sensitive mental health components within existing healthcare programs could be a first measure to alleviate the mental health burden. The ``National Action Plan for Climate Change and Human Health'' of the Indian government is a first step in this direction and aims to strengthen the Indian health care systems' response to climate-sensitive illness and to raise awareness about the impacts of climate change on health \citep{ncdc2021}. A particular focus hereby should be on rural areas; as heatwaves become more frequent and severe, and access to climate adaption measures is often limited in rural areas, policies that provide safety nets might be necessary for vulnerable populations in India.

Moreover, in terms of the scientific sphere, researchers should advocate for the inclusion of wet bulb temperature in climate risk assessments and health impact models. This will ensure more accurate predictions of climate-related health burdens. Also, establishing long-term monitoring systems for mental health outcomes in the context of climate change is essential to generate new insights about the temporal dynamics and causal pathways, particular in LMICs, where data on mental health is still sparse.

Overall, this study contributes to a growing body of evidence that highlights the multifaceted impacts of climate change on human well-being and underscores the need to address the mental health dimensions of climate change. By highlighting the compounded risks posed by heat and humidity, it advocates for integrating mental health into broader climate resilience strategies, in particular in LMICs in South- and Southeast Asia, where high humidity exacerbates the physiological and psychological burdens of extreme heat.

\section*{Declaration of generative AI and AI-assisted technologies in the writing process}
During the preparation of this work the author used ChatGTP in order to shorten the manuscript and to improve the readability and language of the manuscript. After using this tool, the author reviewed and edited the content as needed and takes full responsibility for the content of the published article.

\newpage
\interlinepenalty=10000
\bibliographystyle{apalike}
\bibliography{references}

\begin{thebibliography}{}

\bibitem[{American Psychological Association}, 2025]{apa2025}
{American Psychological Association} (2025).
\newblock {APA Dictionary of Psychology}.
\newblock Last accessed: 22.01.2025. \url{https://dictionary.apa.org/anxiety}.

\bibitem[Arokiasamy et~al., 2013]{sage1}
Arokiasamy, P., Parasuraman, S., Sekher, T.~V., and Lhungdim, H. (2013).
\newblock {Study on global AGEing and adult health (SAGE), Wave 1, 2007}.
\newblock International Institute for Population Sciences and World Health
  Organization. Last accessed: 16.01.2025.
  \url{https://www.iipsindia.ac.in/sites/default/files/SAGE_Wave1_India_Report.pdf}.

\bibitem[Arokiasamy et~al., 2020]{sage2}
Arokiasamy, P., Sekher, T., Lhungdim, H., Dhar, M., , and Roy, A. (2020).
\newblock {Study on global AGEing and adult health (SAGE) Wave 2, India
  National Report}.
\newblock International Institute for Population Sciences and World Health
  Organization. Last accessed: 16.01.2025.
  \url{https://iipsindia.ac.in/sites/default/files/other_files/WHO-SAGE_Wave-2_India_Report.pdf}.

\bibitem[Ballester et~al., 2023]{ballester2023heat}
Ballester, J., Quijal-Zamorano, M., M{\'e}ndez~Turrubiates, R.~F., Pegenaute,
  F., Herrmann, F.~R., Robine, J.~M., Basaga{\~n}a, X., Tonne, C., Ant{\'o},
  J.~M., and Achebak, H. (2023).
\newblock Heat-related mortality in {Europe} during the summer of 2022.
\newblock {\em Nature Medicine}, 29(7):1857--1866.

\bibitem[Banerjee and Maharaj, 2020]{banerjee2020heat}
Banerjee, R. and Maharaj, R. (2020).
\newblock Heat, infant mortality, and adaptation: Evidence from {India}.
\newblock {\em Journal of Development Economics}, 143:102378.

\bibitem[Barreca et~al., 2016]{barreca2016adapting}
Barreca, A., Clay, K., Deschenes, O., Greenstone, M., and Shapiro, J.~S.
  (2016).
\newblock Adapting to climate change: The remarkable decline in the us
  temperature-mortality relationship over the twentieth century.
\newblock {\em Journal of Political Economy}, 124(1):105--159.

\bibitem[Baylis, 2020]{baylis2020temperature}
Baylis, P. (2020).
\newblock Temperature and temperament: Evidence from {Twitter}.
\newblock {\em Journal of Public Economics}, 184:104161.

\bibitem[Beaudoing et~al., 2020]{NASA1}
Beaudoing, H., Rodell, M., and NASA/GSFC/HSL (2020).
\newblock {GLDAS Noah Land Surface Model L4 3 hourly 0.25 x 0.25 degree V2.1}.
\newblock Greenbelt, Maryland: USA. Goddard Earth Sciences Data and Information
  Services Center (GES DISC). Last accessed: 21.11.2023. Doi:
  10.5067/E7TYRXPJKWOQ.
  \url{https://disc.gsfc.nasa.gov/datasets/GLDAS_NOAH025_3H_2.1/summary}.

\bibitem[Beck et~al., 2018]{beck2018present}
Beck, H.~E., Zimmermann, N.~E., McVicar, T.~R., Vergopolan, N., Berg, A., and
  Wood, E.~F. (2018).
\newblock Present and future {K{\"o}ppen-Geiger} climate classification maps at
  1-km resolution.
\newblock {\em Scientific Data}, 5(1):1--12.

\bibitem[Berry et~al., 2010]{berry2010climate}
Berry, H.~L., Bowen, K., and Kjellstrom, T. (2010).
\newblock Climate change and mental health: A causal pathways framework.
\newblock {\em International Journal of Public Health}, 55:123--132.

\bibitem[Berry et~al., 2018]{berry2018case}
Berry, H.~L., Waite, T.~D., Dear, K.~B., Capon, A.~G., and Murray, V. (2018).
\newblock The case for systems thinking about climate change and mental health.
\newblock {\em Nature Climate Change}, 8(4):282--290.

\bibitem[Bharadwaj et~al., 2017]{bharadwaj2017mental}
Bharadwaj, P., Pai, M.~M., and Suziedelyte, A. (2017).
\newblock Mental health stigma.
\newblock {\em Economics Letters}, 159:57--60.

\bibitem[Bubonya et~al., 2017]{bubonya2017mental}
Bubonya, M., Cobb-Clark, D.~A., and Wooden, M. (2017).
\newblock Mental health and productivity at work: Does what you do matter?
\newblock {\em Labour Economics}, 46:150--165.

\bibitem[Burgess et~al., 2017]{burgess2017weather}
Burgess, R., Deschenes, O., Donaldson, D., and Greenstone, M. (2017).
\newblock Weather, climate change and death in {India}.
\newblock Working Paper. Last accessed: 16.01.2025.
  \url{https://epic.uchicago.edu/wp-content/uploads/2019/07/Publication-9.pdf}.

\bibitem[Burke et~al., 2018]{burke2018higher}
Burke, M., Gonz{\'a}lez, F., Baylis, P., Heft-Neal, S., Baysan, C., Basu, S.,
  and Hsiang, S. (2018).
\newblock Higher temperatures increase suicide rates in the {United States and
  Mexico}.
\newblock {\em Nature Climate Change}, 8(8):723--729.

\bibitem[Carleton, 2017]{carleton2017crop}
Carleton, T.~A. (2017).
\newblock Crop-damaging temperatures increase suicide rates in {India}.
\newblock {\em Proceedings of the National Academy of Sciences},
  114(33):8746--8751.

\bibitem[Cattaneo and Peri, 2016]{cattaneo2016migration}
Cattaneo, C. and Peri, G. (2016).
\newblock The migration response to increasing temperatures.
\newblock {\em Journal of Development Economics}, 122:127--146.

\bibitem[Clayton and Karazsia, 2020]{clayton2020development}
Clayton, S. and Karazsia, B.~T. (2020).
\newblock Development and validation of a measure of climate change anxiety.
\newblock {\em Journal of Environmental Psychology}, 69:101434.

\bibitem[Dell et~al., 2012]{dell2012temperature}
Dell, M., Jones, B.~F., and Olken, B.~A. (2012).
\newblock Temperature shocks and economic growth: Evidence from the last half
  century.
\newblock {\em American Economic Journal: Macroeconomics}, 4(3):66--95.

\bibitem[Desch{\^e}nes and Greenstone, 2011]{deschenes2011climate}
Desch{\^e}nes, O. and Greenstone, M. (2011).
\newblock Climate change, mortality, and adaptation: Evidence from annual
  fluctuations in weather in the {US}.
\newblock {\em American Economic Journal: Applied Economics}, 3(4):152--185.

\bibitem[Desch{\^e}nes et~al., 2009]{deschenes2009climate}
Desch{\^e}nes, O., Greenstone, M., and Guryan, J. (2009).
\newblock Climate change and birth weight.
\newblock {\em American Economic Review}, 99(2):211--217.

\bibitem[{Directorate General of Health Services}, 2024]{dghs}
{Directorate General of Health Services} (2024).
\newblock {National Mental Health Programme. Release of Funds to various
  States/UTs under District Mental Health Programme.}
\newblock Last accessed: 13.01.2025.
  \url{https://dghs.gov.in/content/1350_3_NationalMentalHealthProgramme.aspx}.

\bibitem[El~Khayat et~al., 2022]{el2022impacts}
El~Khayat, M., Halwani, D.~A., Hneiny, L., Alameddine, I., Haidar, M.~A., and
  Habib, R.~R. (2022).
\newblock Impacts of climate change and heat stress on farmworkers' health: A
  scoping review.
\newblock {\em Frontiers in Public Health}, 10:782811.

\bibitem[Escobar~Carias et~al., 2024]{carias2024temperature}
Escobar~Carias, M., Johnston, D.~W., Knott, R., and Sweeney, R. (2024).
\newblock Temperature’s toll on decision-making.
\newblock {\em The Economic Journal}, 134(663):2746--2771.

\bibitem[Fritz, 2022]{fritz2022temperature}
Fritz, M. (2022).
\newblock Temperature and non-communicable diseases: Evidence from
  {Indonesia's} primary health care system.
\newblock {\em Health Economics}, 31(11):2445--2464.

\bibitem[Gangadhar et~al., 2023]{gangadhar2023mental}
Gangadhar, B., Kumar, C.~N., Sadh, K., Manjunatha, N., Math, S.~B., Kalaivanan,
  R.~C., Rao, G.~N., Parthasarathy, R., Chand, P.~K., Chandra, P.~S., et~al.
  (2023).
\newblock Mental health programme in {India}: Has the tide really turned?
\newblock {\em Indian Journal of Medical Research}, 157(5):387--394.

\bibitem[Girase et~al., 2022]{girase2022india}
Girase, B., Parikh, R., Vashisht, S., Mullick, A., Ambhore, V., and Maknikar,
  S. (2022).
\newblock India's policy and programmatic response to mental health of young
  people: A narrative review.
\newblock {\em Social Science and Medicine - Mental Health}, 2:100145.

\bibitem[{Global Burden of Disease Collaborative Network}, 2020]{gbd2019}
{Global Burden of Disease Collaborative Network} (2020).
\newblock {Global Burden of Disease Study 2019}.
\newblock Seattle, United States: Institute for Health Metrics and Evaluation.

\bibitem[{Global Modeling and Assimilation Office (GMAO)}, 2015]{MERRA2}
{Global Modeling and Assimilation Office (GMAO)} (2015).
\newblock {MERRA-2 tavg1 2 aer Nx: 2d, 1-Hourly, Time-averaged, Single-Level,
  Assimilation, Aerosol Diagnostics V5.12.4}.
\newblock Greenbelt, Maryland: USA. Goddard Earth Sciences Data and Information
  Services Center (GES DISC). Last accessed: 21.11.2023. Doi:
  10.5067/KLICLTZ8EM9D.
  \url{https://disc.gsfc.nasa.gov/datasets/M2T1NXAER_5.12.4/summary}.

\bibitem[{Global Modeling and Assimilation Office (GMAO)}, 2022]{NASA2022}
{Global Modeling and Assimilation Office (GMAO)} (2022).
\newblock {Modern-Era Retrospective analysis for Research and Applications,
  Version 2. MERRA-2 FAQ}.
\newblock Last accessed: 03.01.2024.
  \url{https://gmao.gsfc.nasa.gov/reanalysis/MERRA-2/FAQ/#Q4}.

\bibitem[Hansen et~al., 2008]{hansen2008effect}
Hansen, A., Bi, P., Nitschke, M., Ryan, P., Pisaniello, D., and Tucker, G.
  (2008).
\newblock The effect of heatwaves on mental health in a temperate {Australian}
  city.
\newblock {\em Epidemiology}, 19(6):S85.

\bibitem[Heyes and Saberian, 2019]{heyes2019temperature}
Heyes, A. and Saberian, S. (2019).
\newblock Temperature and decisions: Evidence from 207,000 court cases.
\newblock {\em American Economic Journal: Applied Economics}, 11(2):238--265.

\bibitem[Heyes and Saberian, 2022]{heyes2022hot}
Heyes, A. and Saberian, S. (2022).
\newblock Hot days, the ability to work and climate resilience: Evidence from a
  representative sample of 42,152 {Indian} households.
\newblock {\em Journal of Development Economics}, 155:102786.

\bibitem[Hsiang et~al., 2013]{hsiang2013quantifying}
Hsiang, S.~M., Burke, M., and Miguel, E. (2013).
\newblock Quantifying the influence of climate on human conflict.
\newblock {\em Science}, 341(6151):1235367.

\bibitem[Hua et~al., 2023]{hua2023effects}
Hua, Y., Qiu, Y., and Tan, X. (2023).
\newblock {The effects of temperature on mental health: Evidence from China}.
\newblock {\em Journal of Population Economics}, 36(3):1293--1332.

\bibitem[Jayasankar et~al., 2022]{jayasankar2022epidemiology}
Jayasankar, P., Manjunatha, N., Rao, G.~N., Gururaj, G., Varghese, M., Benegal,
  V., et~al. (2022).
\newblock Epidemiology of common mental disorders: Results from {``National
  Mental Health Survey'' of India}, 2016.
\newblock {\em Indian Journal of Psychiatry}, 64(1):13--19.

\bibitem[Karlsson and Ziebarth, 2018]{karlsson2018population}
Karlsson, M. and Ziebarth, N.~R. (2018).
\newblock Population health effects and health-related costs of extreme
  temperatures: Comprehensive evidence from {Germany}.
\newblock {\em Journal of Environmental Economics and Management}, 91:93--117.

\bibitem[Kaur et~al., 2021]{kaur2021systematic}
Kaur, A., Kallakuri, S., Kohrt, B.~A., Heim, E., Gronholm, P.~C., Thornicroft,
  G., and Maulik, P.~K. (2021).
\newblock Systematic review of interventions to reduce mental health stigma in
  {India}.
\newblock {\em Asian Journal of Psychiatry}, 55:102466.

\bibitem[Liu et~al., 2021]{liu2021there}
Liu, J., Varghese, B.~M., Hansen, A., Xiang, J., Zhang, Y., Dear, K., Gourley,
  M., Driscoll, T., Morgan, G., Capon, A., et~al. (2021).
\newblock Is there an association between hot weather and poor mental health
  outcomes? {A systematic review and meta-analysis}.
\newblock {\em Environment International}, 153:106533.

\bibitem[Liu et~al., 2019]{liu2019influence}
Liu, X., Liu, H., Fan, H., Liu, Y., and Ding, G. (2019).
\newblock Influence of heat waves on daily hospital visits for mental illness
  in {Jinan, China}—a case-crossover study.
\newblock {\em International Journal of Environmental Research and Public
  Health}, 16(1):87.

\bibitem[LoPalo, 2023]{lopalo2023temperature}
LoPalo, M. (2023).
\newblock Temperature, worker productivity, and adaptation: Evidence from
  survey data production.
\newblock {\em American Economic Journal: Applied Economics}, 15(1):192--229.

\bibitem[LoPalo et~al., 2019]{lopalo2019quantifying}
LoPalo, M., Kuruc, K., Budolfson, M., and Spears, D. (2019).
\newblock Quantifying {India's} climate vulnerability.
\newblock Policy Report for the Indian Policy Forum, Vol 15. 107--148.

\bibitem[Luan et~al., 2019]{luan2019associations}
Luan, G., Yin, P., Wang, L., and Zhou, M. (2019).
\newblock Associations between ambient high temperatures and suicide mortality:
  A multi-city time-series study in {China}.
\newblock {\em Environmental Science and Pollution Research}, 26:20377--20385.

\bibitem[Mani et~al., 2013]{mani2013poverty}
Mani, A., Mullainathan, S., Shafir, E., and Zhao, J. (2013).
\newblock Poverty impedes cognitive function.
\newblock {\em Science}, 341(6149):976--980.

\bibitem[Matthews et~al., 2017]{matthews2017communicating}
Matthews, T.~K., Wilby, R.~L., and Murphy, C. (2017).
\newblock Communicating the deadly consequences of global warming for human
  heat stress.
\newblock {\em Proceedings of the National Academy of Sciences},
  114(15):3861--3866.

\bibitem[{Ministry of Health \& Family Welfare}, 2024]{mohfw}
{Ministry of Health \& Family Welfare} (2024).
\newblock {National Mental Health Programme (NMHP). List of Approved DMHP
  Districts}.
\newblock Last accessed: 08.01.2025.
  \url{https://mohfw.gov.in/sites/default/files/List%20of%20Approved%20DMHP%20Districts_0.pdf}.

\bibitem[Mukherjee and Sanders, 2021]{mukherjee2021causal}
Mukherjee, A. and Sanders, N.~J. (2021).
\newblock The causal effect of heat on violence: Social implications of
  unmitigated heat among the incarcerated.
\newblock NBER Working Paper 28987. National Bureau of Economic Research.

\bibitem[Mullins and White, 2019]{mullins2019temperature}
Mullins, J.~T. and White, C. (2019).
\newblock Temperature and mental health: Evidence from the spectrum of mental
  health outcomes.
\newblock {\em Journal of Health Economics}, 68:102240.

\bibitem[{National Centre for Disease Control (India)}, 2021]{ncdc2021}
{National Centre for Disease Control (India)} (2021).
\newblock {National Action Plan on Climate Change and Human Health.}
\newblock Last accessed: 14.01.2025.
  \url{https://ncdc.mohfw.gov.in/wp-content/uploads/2024/07/NATIONAL-ACTION-PLAN-FOR-CLIMATE-CHANGE-HUMAN-HEALTH-2021.pdf}.

\bibitem[Noelke et~al., 2016]{noelke2016increasing}
Noelke, C., McGovern, M., Corsi, D.~J., Jimenez, M.~P., Stern, A., Wing, I.~S.,
  and Berkman, L. (2016).
\newblock Increasing ambient temperature reduces emotional well-being.
\newblock {\em Environmental Research}, 151:124--129.

\bibitem[Obradovich et~al., 2017]{obradovich2017nighttime}
Obradovich, N., Migliorini, R., Mednick, S.~C., and Fowler, J.~H. (2017).
\newblock Nighttime temperature and human sleep loss in a changing climate.
\newblock {\em Science Advances}, 3(5):e1601555.

\bibitem[Obradovich et~al., 2018]{obradovich2018empirical}
Obradovich, N., Migliorini, R., Paulus, M.~P., and Rahwan, I. (2018).
\newblock Empirical evidence of mental health risks posed by climate change.
\newblock {\em Proceedings of the National Academy of Sciences},
  115(43):10953--10958.

\bibitem[Pai et~al., 2014]{pai2014development}
Pai, D., Rajeevan, M., Sreejith, O., Mukhopadhyay, B., and Satbha, N. (2014).
\newblock Development of a new high spatial resolution (0.25$\times$0.25) long
  period (1901-2010) daily gridded rainfall data set over {India} and its
  comparison with existing data sets over the region.
\newblock {\em Mausam}, 65(1):1--18.

\bibitem[Pailler and Tsaneva, 2018]{pailler2018effects}
Pailler, S. and Tsaneva, M. (2018).
\newblock The effects of climate variability on psychological well-being in
  {India}.
\newblock {\em World Development}, 106:15--26.

\bibitem[Raymond et~al., 2020]{raymond2020emergence}
Raymond, C., Matthews, T., and Horton, R.~M. (2020).
\newblock The emergence of heat and humidity too severe for human tolerance.
\newblock {\em Science Advances}, 6(19):1838.

\bibitem[Ridley et~al., 2020]{ridley2020poverty}
Ridley, M., Rao, G., Schilbach, F., and Patel, V. (2020).
\newblock Poverty, depression, and anxiety: Causal evidence and mechanisms.
\newblock {\em Science}, 370(6522):0214.

\bibitem[Saeed et~al., 2021]{saeed2021deadly}
Saeed, F., Schleussner, C.-F., and Ashfaq, M. (2021).
\newblock Deadly heat stress to become commonplace across {South Asia} already
  at 1.5 degree {C} of global warming.
\newblock {\em Geophysical Research Letters}, 48(7):e2020GL091191.

\bibitem[Sagar et~al., 2020]{sagar2020burden}
Sagar, R., Dandona, R., Gururaj, G., Dhaliwal, R., Singh, A., Ferrari, A., Dua,
  T., Ganguli, A., Varghese, M., Chakma, J.~K., et~al. (2020).
\newblock {The burden of mental disorders across the states of India: the
  Global Burden of Disease Study 1990--2017}.
\newblock {\em The Lancet Psychiatry}, 7(2):148--161.

\bibitem[Sanz-Barbero et~al., 2018]{sanz2018heat}
Sanz-Barbero, B., Linares, C., Vives-Cases, C., Gonz{\'a}lez, J.~L.,
  L{\'o}pez-Ossorio, J.~J., and D{\'\i}az, J. (2018).
\newblock Heat wave and the risk of intimate partner violence.
\newblock {\em Science of the Total Environment}, 644:413--419.

\bibitem[Stull, 2011]{stull2011wet}
Stull, R. (2011).
\newblock Wet-bulb temperature from relative humidity and air temperature.
\newblock {\em Journal of Applied Meteorology and Climatology},
  50(11):2267--2269.

\bibitem[Thompson et~al., 2018]{thompson2018associations}
Thompson, R., Hornigold, R., Page, L., and Waite, T. (2018).
\newblock Associations between high ambient temperatures and heat waves with
  mental health outcomes: A systematic review.
\newblock {\em Public Health}, 161:171--191.

\bibitem[Trang et~al., 2016]{trang2016heatwaves}
Trang, P.~M., Rockl{\"o}v, J., Giang, K.~B., Kullgren, G., and Nilsson, M.
  (2016).
\newblock Heatwaves and hospital admissions for mental disorders in northern
  {Vietnam}.
\newblock {\em PloS One}, 11(5):e0155609.

\bibitem[{United Nations Statistics Division}, 2025]{UNdata2025}
{United Nations Statistics Division} (2025).
\newblock {Life expectancy at birth for both sexes combined (years) - India}.
\newblock Last accessed: 23.01.2025.
  \url{https://data.un.org/Data.aspx?q=india+life+expectancy+1965&d=PopDiv&f=variableID%3A68%3BcrID%3A356%3BtimeID%3A103%2C104}.

\bibitem[Wang et~al., 2018]{wang2018effect}
Wang, S., Zhang, X., Xie, M., Zhao, D., Zhang, H., Zhang, Y., Cheng, Q., Bai,
  L., and Su, H. (2018).
\newblock Effect of increasing temperature on daily hospital admissions for
  schizophrenia in {Hefei, China}: a time-series analysis.
\newblock {\em Public Health}, 159:70--77.

\bibitem[White, 2017]{white2017dynamic}
White, C. (2017).
\newblock The dynamic relationship between temperature and morbidity.
\newblock {\em Journal of the Association of Environmental and Resource
  Economists}, 4(4):1155--1198.

\bibitem[Wood et~al., 2019]{wood2019daily}
Wood, W.~H., Marshall, S.~J., and Fargey, S.~E. (2019).
\newblock Daily measurements of near-surface humidity from a mesonet in the
  foothills of the canadian rocky mountains, 2005--2010.
\newblock {\em Earth System Science Data}, 11(1):23--34.

\bibitem[{World Bank}, 2025]{worldbank2025}
{World Bank} (2025).
\newblock {Survival to age 65, female (\% of cohort) - India}.
\newblock Last accessed: 23.01.2025.
  \url{https://data.worldbank.org/indicator/SP.DYN.TO65.FE.ZS?locations=IN}.

\bibitem[{World Meteorological Organization (WMO)}, 2008]{WMO2008}
{World Meteorological Organization (WMO)} (2008).
\newblock Guide to meteorological instruments and methods of observation.
\newblock Geneva: Switzerland. Secretariat of the World Meteorological
  Organization. Last accessed: 03.01.2024.
  \url{https://www.seedmech.com/documents_folder/wmo_no_8.pdf}.

\bibitem[Wu et~al., 2020]{wu2020climate}
Wu, J., Snell, G., and Samji, H. (2020).
\newblock Climate anxiety in young people: A call to action.
\newblock {\em The Lancet Planetary Health}, 4(10):e435--e436.

\bibitem[Xue et~al., 2019]{xue2019declines}
Xue, T., Zhu, T., Zheng, Y., and Zhang, Q. (2019).
\newblock Declines in mental health associated with air pollution and
  temperature variability in {China}.
\newblock {\em Nature Communications}, 10(1):2165.

\bibitem[Zhang et~al., 2023]{zhang2023heatwave}
Zhang, X., Chen, F., and Chen, Z. (2023).
\newblock Heatwave and mental health.
\newblock {\em Journal of Environmental Management}, 332:117385.

\bibitem[Zivin et~al., 2020]{zivin2020temperature}
Zivin, J.~G., Song, Y., Tang, Q., and Zhang, P. (2020).
\newblock Temperature and high-stakes cognitive performance: Evidence from the
  national college entrance examination in china.
\newblock {\em Journal of Environmental Economics and Management}, 104:102365.

\end{thebibliography}

\newpage
\appendix
\pagenumbering{arabic}

\section*{Appendices}
\subsection*{Appendix A: Calculations}\label{AppendixA}
% !TEX spellcheck = en_US
\textit{\textbf{Calculation of wet bulb temperature}}
\setcounter{equation}{0}

\noindent
Wet bulb temperature is a non-linear function of the dry bulb temperature and varies with the level of relative humidity. I follow  \cite{WMO2008} and  \cite{wood2019daily} to derive values for relative humidity and wet bulb temperature. 
The NASA data contains values only for specific humidity ($q$), from which, together with the values of temperature ($t$) and pressure ($p$), relative humidity can be calculated. 

\noindent
\textbf{Step 1}: Calculation of relative humidity

\noindent
Relative humidity can be expressed as the mass of water vapor per unit mass of dry air (the mixing ratio) relative to the saturation mixing ratio

\begin{equation}
	RH = 100*\frac{w}{w'}
\end{equation}

\noindent
where $w$ is commonly assumed to be equal to specific humidity $q$ \citep{wood2019daily}, hence

\begin{equation}
	RH = 100*\frac{q}{w'}
\end{equation}

\noindent
The saturation mixing ratio $w'$, in turn depends on a given air pressure $p$ and the saturation vapor pressure $e'$ in the following relationship

\begin{equation}
	w' = 0.622* \frac{e'}{p}
\end{equation}
\noindent
where $e'$ (in PA) is a function of temperature (in in °C):

\begin{equation}
	e' =  611 exp  \left(  \frac{ 17.62t }{ 243.12 + t} \right) 
\end{equation}

\noindent
leading to the following equation in which specific humidity ($q$, in kg/kg), pressure ($p$, in Pa) and temperature $t$ (in in °C) can be directly used to calculate relative humidity. 

\begin{equation}
RH = 0.263 * p * q *  \left[ exp  \left(  \frac{ 17.62t }{ 243.12+t} \right) \right]^{-1}
\end{equation}

\noindent
\textbf{Step 2}: Calculation of wet bulb temperature using dry bulb temperature (t, in °C) and relative humidity (RH, in \%). Calculations follow \cite{stull2011wet} and \cite{lopalo2023temperature}.

\begin{multline}
wbulb=t * [ atan(0.151997 * (RH+8.313658)^{0.5}] + atan(t+RH) - atan(RH-1.676331) \\
+ 0.00391838 * RH^{3/2} * atan(0.023101*RH) - 4.686035
\end{multline}

\newpage
\noindent
\textit{\textbf{Calculation of air pollution (PM2.5)}}

\noindent
Information on air pollution is drawn from the NASA MERRA-2 data set \citep{MERRA2}. This data set contains information on black carbon column mass density (BCSMASS), dust column mass density (DUSMASS25), organic carbon column mass density (OCSMASS), sea salt surface mass concentration (SSSMASS25), and SO4 surface mass concentration (SO4SMASS), which are combined to PM2.5 using the following formula provided by the NASA \citep{NASA2022}:
\begin{multline}
	PM2.5 = DUSMASS25 + OCSMASS+ BCSMASS + SSSMASS25 + \\ SO4SMASS* (132.14/96.06)
\end{multline}

\clearpage

\subsection*{Appendix B: Summary statistics - Dry bulb temperature}
% !TEX spellcheck = en_US
\setcounter{table}{0} 
\renewcommand{\thetable}{B\arabic{table}}
\vspace*{\fill}
\begin{table}[h!]
	\caption{\label{sumstats_dry}Summary statistics - Dry bulb temperature}
	\small
	\centering
	\begin{tabular}{lrrrr}
		\midrule
		\midrule 
                                 & Mean & SD   & Min & Max \\
                                 \midrule
    Dry bulb temperature &       29.87&        4.52&          11.42&          42.62\\
    & & & & \\
\# Days \textless 23.5°C&        3.87&        8.18&           0&          30\\
\# Days 23.5°C-25°C&        1.73&        3.43&           0&          24\\
\# Days 25°C-26.5°C&        2.20&        3.71&           0&          30\\
\# Days 26.5°C-28°C&        2.87&        3.87&           0&          28\\
\# Days 28°C-29.5°C&        3.74&        5.05&           0&          23\\
\# Days 29.5°C-31°C&        3.43&        4.43&           0&          25\\
\# Days 31°C-32.5°C&        3.11&        4.31&           0&          21\\
\# Days 32.5°C-34°C&        2.71&        3.77&           0&          19\\
\# Days 34°C-35.5°C&        2.62&        4.11&           0&          19\\
\# Days 35.5°C-37°C&        2.09&        3.79&           0&          19\\
\# Days $\geq$ 37°C&        1.62&        4.12&           0&          28\\
    & & & & \\
Heat wave ( days $\geq$ 37°)&        0.16&        0.36&           0&           1\\
Heat wave ($\geq$ 3 days $\geq$ 37°)&        0.14&        0.34&           0&           1\\
\# Consecutive days $\geq$ 37°&        1.11&        3.10&           0&          26\\
    & & & & \\
\# Days \textless 25.5°C&        6.28&       10.25&           0&          30\\
\# Days 25.5.5°C-27°C&        2.39&        3.77&           0&          30\\
\# Days 27°C-28.5°C&        3.15&        4.17&           0&          28\\
\# Days 28.5°C-30°C&        3.86&        5.18&           0&          26\\
\# Days 30°C-31.5°C&        3.19&        4.03&           0&          25\\
\# Days 31.5°C-33°C&        3.06&        4.09&           0&          20\\
\# Days 33°C-34.5°C&        2.67&        3.96&           0&          20\\
\# Days 34.5°C-36°C&        2.48&        4.01&           0&          19\\
\# Days 36°C-37.5°C&        1.82&        3.61&           0&          20\\
\# Days 37.5°C-39°C&        0.78&        2.07&           0&          15\\
\# Days $\geq$ 39°C&        0.32&        1.55&           0&          20\\
    & & & & \\
Heat wave ($\geq$ 2 days $\geq$ 39°)&        0.05&        0.21&           0&           1\\
Heat wave ($\geq$ 3 days $\geq$ 39°)&        0.04&        0.19&           0&           1\\
\# Consecutive days $\geq$ 39°&        0.23&        1.04&           0&          13\\
\midrule
Observations        &       30,300&            &            &            \\
\midrule
\midrule
	\multicolumn{5}{p{8cm}}{}   

	\end{tabular}
\end{table}
\vspace*{\fill}

\clearpage

\subsection*{Appendix C: Results - Additional tables}
\setcounter{table}{0} 
\renewcommand{\thetable}{C\arabic{table}}

\setcounter{figure}{0} 
\renewcommand{\thefigure}{C\arabic{figure}}

% !TEX spellcheck = en_US

 \vspace*{\fill}
\begin{table}[h!]
	\caption{\label{wetbulb_R1}Wet bulb temperature and mental health - Main results}
	\small
	\centering
	\begin{tabular}{lcccc}
		\midrule
		\midrule
                  & (1)              & (2)                               & (3)                 & (4)                                    \\
& Depression  & Severe/Extreme  & Worry/Anxiety  & Severe/Extreme  \\
&  Score &  Depression (=1) &  Score &  Anxiety/Worry   (=1) \\
		\midrule
&                  &                                   &                     &                                        \\
\# Days \textless 16.5°C  & -0.012** &	-0.105&	-0.005&	0.049
                                \\
&(0.005)&	(0.112)	&(0.005)&	(0.130)
                            \\
\# Days 16.5°C-18°C       & -0.007*&	-0.017&	-0.004&	0.188
                                 \\
& (0.004)&	(0.110)&	(0.005)&	(0.130)
                               \\
\# Days 18°C-19.5°C       &-0.009**	&-0.116	&-0.007	&-0.135
                                \\
&(0.004)&	(0.104)&	(0.005)&	(0.128)
                               \\
\# Days 19.5°C-21°C       &-0.000&	0.094&	-0.002&	0.236*
                                \\
& (0.004)&	(0.106)&	(0.005)&(0.123)
                             \\
\# Days 21°-22.5°C       & Ref.           & Ref.                              & Ref.               & Ref.                                \\
& & & & \\
\# Days 22.5°C-24°C       &0.001&	0.156*&	-0.002&	0.191*
                                \\
&(0.004)&	(0.093)&	(0.004)	&(0.108)
                              \\
\# Days 24°C-25.5°C       &0.004&	0.117	&-0.002&	0.137
                                \\
& (0.003)&	(0.086)	&(0.004)&	(0.106)
                              \\
\# Days 25.5°C-27°C       &0.008*&	0.258**	&-0.002	&0.128
                                 \\
&(0.004)&	(0.105)	&(0.005)&	(0.128)
                              \\
\# Days \textgreater 27°C & 0.017**	&0.497***&	0.005&	0.382*
                                \\
& (0.007)&	(0.162)&	(0.008)	&(0.231)
                             \\
&                  &                                   &                     &                                        \\
Observations      &27,287&	27,287	&27,243&	27,243
                               \\
R-squared         &0.159&	0.070&	0.176&	0.086
                                \\
Weather Controls  & \Checkmark              & \Checkmark                               & \Checkmark                 & \Checkmark                                    \\
Year-month FE   & \Checkmark              & \Checkmark                               & \Checkmark                 & \Checkmark                                    \\
State FE       & \Checkmark              & \Checkmark                               & \Checkmark                 & \Checkmark                                    \\
KGC FE            & \Checkmark              & \Checkmark                               & \Checkmark                 & \Checkmark                                    \\
\# of PSU         & 371              & 371                               & 371                 & 371                                    \\
Mean of dependent variable                 & 1.82             & 8.47\%                               & 2.00                & 12.69\%         \\                         
		\midrule
		\midrule
		\multicolumn{5}{p{16cm}}{\textit{Notes:} The binary outcome variables (Columns (2) and (4)) are multiplied by 100, hence, the coefficients are directly interpretable as percentage points. The depression and anxiety scores ranges from 1-5. Robust standard errors are clustered at the PSU level and presented in parentheses. ***   p\textless{}0.01, ** p\textless{}0.05, * p\textless{}0.1.}
	\end{tabular}
\end{table}
 \vspace*{\fill}

\begin{table}[ht!]
	\caption{\label{wetbulb_R2}Wet bulb temperature and mental health - Different bins (I)}
	\small
	\centering
	\begin{tabular}{lcccc}
		\midrule
		\midrule
		& (1)              & (2)                               & (3)                 & (4)                                    \\
		& Depression  & Severe/Extreme  & Worry/Anxiety  & Severe/Extreme  \\
		&  Score &  Depression (=1) &  Score &  Anxiety/Worry   (=1) \\
		\midrule
		&                  &                                   &                     &                                        \\
\# Days \textless 17.5°C  & -0.010** & -0.186*  & -0.003  & 0.015   \\
& (0.004)  & (0.110)  & (0.005) & (0.126) \\
\# Days 17.5°C-19°C       & -0.007*  & -0.162   & -0.006  & -0.220* \\
& (0.004)  & (0.113)  & (0.004) & (0.128) \\
\# Days 19°C-20.5°C       & -0.002   & -0.047   & -0.002  & 0.089   \\
& (0.004)  & (0.112)  & (0.004) & (0.120) \\
\# Days 20.5°C-22°C       & -0.000   & -0.146   & 0.000   & -0.140  \\
& (0.003)  & (0.091)  & (0.004) & (0.099) \\
\# Days 22°C-23.5°C       & Ref.     & Ref.     & Ref.    & Ref.    \\
&          &          &         &         \\
\# Days 23.5°C-25°C       & 0.003    & -0.017   & -0.002  & -0.053  \\
& (0.003)  & (0.083)  & (0.004) & (0.101) \\
\# Days 25°C-26.5°C       & 0.006*   & 0.049    & -0.001  & -0.050  \\
& (0.003)  & (0.081)  & (0.004) & (0.099) \\
\# Days 26.5°C-28°C       & 0.005    & 0.100    & -0.005  & -0.197  \\
& (0.005)  & (0.125)  & (0.007) & (0.180) \\
\# Days \textgreater 28°C & 0.113*** & 2.452*** & 0.095** & 3.109** \\
& (0.027)  & (0.648)  & (0.040) & (1.227) \\
&          &          &         &         \\
		Observations      & 27,287   & 27,287   & 27,243  & 27,243  \\
		R-squared         & 0.160    & 0.071    & 0.177   & 0.087  \\
		Weather Controls  & \Checkmark              & \Checkmark                               & \Checkmark                 & \Checkmark                                    \\
		Year-month FE   & \Checkmark              & \Checkmark                               & \Checkmark                 & \Checkmark                                    \\
		State FE       & \Checkmark              & \Checkmark                               & \Checkmark                 & \Checkmark                                    \\
		KGC FE            & \Checkmark              & \Checkmark                               & \Checkmark                 & \Checkmark                                    \\
		\# of PSU         & 371              & 371                               & 371                 & 371                                    \\
		Mean of dependent variable                & 1.82             & 8.47\%                               & 2.00                & 12.69\%         \\                         
		\midrule
		\midrule
		\multicolumn{5}{p{16cm}}{\textit{Notes:} The binary outcome variables (Columns (2) and (4)) are multiplied by 100, hence, the coefficients are directly interpretable as percentage points. The depression and anxiety scores ranges from 1-5. Robust standard  errors are clustered at the PSU level and presented in parentheses. ***   p\textless{}0.01, ** p\textless{}0.05, * p\textless{}0.1.}
	\end{tabular}
\end{table}

\begin{table}[ht!]
	\caption{\label{wetbulb_R3}Wet bulb temperature and mental health - Different bins (II)}
	\small
	\centering
	\begin{tabular}{lcccc}
		\midrule
		\midrule
		& (1)              & (2)                               & (3)                 & (4)                                    \\
		& Depression  & Severe/Extreme  & Worry/Anxiety  & Severe/Extreme  \\
		&  Score &  Depression (=1) &  Score &  Anxiety/Worry   (=1) \\
		\midrule
		&                  &                                   &                     &                                        \\
\# Days \textless 15.5°C  & -0.012** & -0.031   & -0.006   & 0.038   \\
& (0.005)  & (0.128)  & (0.006)  & (0.145) \\
\# Days 15.5°C-17°C & -0.007   & -0.096   & 0.004    & 0.195   \\
& (0.005)  & (0.129)  & (0.006)  & (0.139) \\
\# Days 17°C-18.5°C & -0.010** & -0.075   & -0.011** & -0.206  \\
& (0.005)  & (0.119)  & (0.006)  & (0.151) \\
\# Days 18.5°C-20°C &-0.005   & 0.047    & -0.003   & 0.093   \\
& (0.004)  & (0.113)  & (0.005)  & (0.140) \\
\# Days 20°C-21.5°C       & Ref.     & Ref.     & Ref.    & Ref.    \\
&          &          &         &         \\
\# Days  21.5°C-23°C & -0.000   & 0.062    & 0.001    & 0.003   \\
 & (0.004)  & (0.087)  & (0.004)  & (0.101) \\
\# Days 23°C-24.5°C &0.003    & 0.144*   & -0.002   & 0.047   \\
& (0.003)  & (0.085)  & (0.004)  & (0.105) \\
\# Days 24.5°C-26°C &0.004    & 0.087    & -0.002   & -0.003  \\
& (0.004)  & (0.091)  & (0.005)  & (0.120) \\
\# Days \textgreater 26°C & 0.011**  & 0.327*** & 0.001    & 0.058   \\
& (0.004)  & (0.113)  & (0.005)  & (0.141) \\
		&          &          &         &         \\
		Observations    & 27,287   & 27,287   & 27,243   & 27,243  \\
		R-squared         & 0.158    & 0.070    & 0.176    & 0.085     \\
		Weather Controls  & \Checkmark              & \Checkmark                               & \Checkmark                 & \Checkmark                                    \\
		Year-month FE   & \Checkmark              & \Checkmark                               & \Checkmark                 & \Checkmark                                    \\
		State FE       & \Checkmark              & \Checkmark                               & \Checkmark                 & \Checkmark                                    \\
		KGC FE            & \Checkmark              & \Checkmark                               & \Checkmark                 & \Checkmark                                    \\
		\# of PSU         & 371              & 371                               & 371                 & 371                                    \\
		Mean of dependent variable                 & 1.82             & 8.47\%                               & 2.00                & 12.69\%         \\                         
		\midrule
		\midrule
		\multicolumn{5}{p{16cm}}{\textit{Notes:} The binary outcome variables (Columns (2) and (4)) are multiplied by 100, hence, the coefficients are directly interpretable as percentage points. The depression and anxiety scores ranges from 1-5. Robust standard  errors are clustered at the PSU level and presented in parentheses. ***   p\textless{}0.01, ** p\textless{}0.05, * p\textless{}0.1.}
	\end{tabular}
\end{table}

\begin{table}[]
	\caption{\label{results_anxiety}Alternative heat indicators - Anxiety}
	\small
	\centering
	\begin{tabular}{lccccc}
		\midrule
		\midrule
			& (1)	& (2)	& (3)	& (4)	& (5) \\ 
		& Sev./Extr.	& Sev./Extr. 	& Sev./Extr. 	& Sev./Extr.	& Sev./Extr. \\ 
		& Anxiety 	& Anxiety 	& Anxiety 	& Anxiety 	& Anxiety \\
		\midrule
		\# Days $\geq$ 27°C (w/o other bins)   & 0.192   &         &         &         &         \\
		& (0.183) &         &         &         &         \\
		\# Consecutive days $\geq$27°C   &         & 0.404   &         &         &         \\
		&         & (0.345) &         &         &         \\
		Heatwave ($\geq$2 days $\geq$27°C)&         &         & 0.481   &         &         \\
		&         &         & (1.297) &         &         \\
		Heatwave ($\geq$3 days $\geq$27°C) &         &         &         & 1.434   &         \\
		&         &         &         & (1.614) &         \\
		Heatwave ($\geq$2 days $\geq$28°C)         &         &         &         &         & 4.890  \\
		&         &         &         &         & (3.926) \\
		&         &         &         &         &         \\
		Observations         & 27,243 & 27,243  & 27,243  & 27,243  & 27,243  \\
		R-squared            & 0.085   & 0.085   & 0.085   & 0.085   & 0.085    \\
		Weather Controls     & \Checkmark      & \Checkmark       & \Checkmark       & \Checkmark       & \Checkmark        \\
		Ind. Controls     & \Checkmark      & \Checkmark       & \Checkmark       & \Checkmark       & \Checkmark        \\
		Year-month FE    & \Checkmark        & \Checkmark       & \Checkmark       & \Checkmark       & \Checkmark        \\
		State FE    & \Checkmark        & \Checkmark       & \Checkmark       & \Checkmark       & \Checkmark        \\
		KGC FE               & \Checkmark        & \Checkmark       & \Checkmark       & \Checkmark       & \Checkmark        \\
		\# of PSU            & 371      & 371     & 371     & 371     & 371     \\
		Mean of dependent variable                & 12.69\%             & 12.69\%                              & 12.69\%                & 12.69\%     & 12.69\%       \\

		\midrule
		\midrule
		\multicolumn{6}{p{15cm}}{\textit{Notes:} The outcome variable is a binary indicator and multiplied by 100, hence, the coefficients are directly interpretable as percentage points. Robust standard  errors are clustered at the PSU level and presented in parentheses. ***   p\textless{}0.01, ** p\textless{}0.05, * p\textless{}0.1.}
	\end{tabular}
\end{table}

\pagebreak

\clearpage
% !TEX spellcheck = en_US
\begin{landscape}
\subsection*{Appendix D: Robustness, dynamics and heterogeneity}
\setcounter{table}{0} 
\renewcommand{\thetable}{D\arabic{table}}

 \vspace*{\fill}
\begin{table}[h!]
	\caption{\label{robustness}Robustness}
	\footnotesize
	\centering
	\begin{tabular}{lcccccccc}
\midrule
\midrule
 & (1) & (2) & (3) & (4) & (5) & (6) & (7) & (8)  \\
 & Severe/Extreme & Severe/Extreme& Severe/Extreme & Severe/Extreme & Severe/Extreme. & Severe/Extreme& Severe/Extreme. & Severe/Extreme \\ 
& Depression & Depression & Depression & Depression &Depression & Depression & Depression &  Depression  \\
& \textit{Baseline} & \textit{District FE} & \textit{Year FE + month FE} & \textit{No state FE} & \textit{No KGC FE} & \textit{No weather controls} & \textit{No ind. controls} & \textit{Logit} \\ \hline
&  &  &  &  &  &  &  &    \\
\# Days $\geq$ 27°C & 0.497*** & 0.259* & 0.480*** & 0.644*** & 0.477*** & 0.226* & 0.449*** & 0.062***   \\
& (0.162) & (0.156) & (0.166) & (0.151) & (0.165) & (0.131) & (0.161) & (0.021) \\
	Heatwave ($\geq$2 days $\geq$27°C)  & 1.994** & 2.176** & 2.116** & 4.026*** & 2.165** & 0.742 & 1.813* & 0.412*** \\
& (0.931) & (1.022) & (0.941) & (0.881) & (0.949) & (0.937) & (0.951) & (0.143) \\
&  &  &  &  &  &  &  &  \\
Observations & 27,287 & 27,282 & 27,287 & 27,287 & 27,287 & 27,287 & 29,541 & 27,287  \\
Weather Controls & \Checkmark & \Checkmark & \Checkmark & \Checkmark & \Checkmark & 		 & \Checkmark & \Checkmark \\
Ind. Controls 	& \Checkmark & \Checkmark & \Checkmark 	& \Checkmark & \Checkmark & \Checkmark &  		& \Checkmark \\
Year-month FE 	& \Checkmark & \Checkmark &  			&  		& 		 & \Checkmark 			& \Checkmark	 & \Checkmark  \\
Year FE 		&  			&  			  & \Checkmark 	& 	\Checkmark		 & \Checkmark 		&  			&  		&   \\
Month FE 		&  			&  				& \Checkmark &  \Checkmark		& \Checkmark		& 			 & 		 &   \\
State FE 		& \Checkmark & 			  & \Checkmark & 			 & \Checkmark 		& \Checkmark 			& \Checkmark 		& \Checkmark  \\
District FE		& 			 & \Checkmark & 			 &  		&  		&  			&  		&   \\
KGC FE 			& \Checkmark & \Checkmark & \Checkmark & \Checkmark 		& 		 & \Checkmark 			& \Checkmark 		& \Checkmark  \\
\midrule
\midrule
\multicolumn{9}{p{26cm}}{\textit{Notes:} Robust standard  errors are clustered at the PSU level and presented in parentheses. The outcome variable is binary and multiplied by 100, hence, coefficients are directly interpretable as percentage points. Column (1) shows the baseline specification. Column (2) includes district fixed-effects (Admin 2) instead of state fixed-effects (Admin 1). Column (3) includes year- and months fixed-effects separately instead of month-year fixed effects. In Column (4), no state- or district fixed-effects are included. In Column (5), no Koeppen-Keiger climate zone fixed-effects are included. In Column (6), no weather controls are included. In Column (7), no individual or household controls are included. In Column (8), the specification is a logit specification instead of a linear probability model. ***   p\textless{}0.01, ** p\textless{}0.05, * p\textless{}0.1.}
\end{tabular}
\end{table}
 \vspace*{\fill}

 \vspace*{\fill}
\begin{table}[ht!]
	\caption{\label{wetbulb_dynamics}Dynamic effects - different time horizons}
	\small
	\centering
	\begin{tabular}{L{3.8cm}cccc}
		\midrule
		\midrule
				& (1)          & (2)                 & (3)        & (4)                   \\
		 			&  Severe/Extreme  & Severe/Extreme    & Severe/Extreme      & Severe/Extreme    \\
				&  Depression  &  Depression    &   Depression &  Depression \\
				& ($Days_{t-1}-Days_{t-14}$) & ($Days_{t-15}-Days_{t-30}$) & ($Days_{t-31}-Days_{t-60}$) & ($Days_{t+1}-Days_{t+30}$) \\
		\midrule
		&              &                     &            &                       \\
\# Days \textless   16.5°C & 0.152   & -0.179   & 0.002   & 0.014    \\
& (0.161) & (0.161)  & (0.116) & (0.131)  \\
\# Days 16.5°C-18°C                       & 0.267   & -0.249   & -0.159  & 0.070    \\
& (0.185) & (0.157)  & (0.103) & (0.137)  \\
\# Days 18°C-19.5°C                       & -0.105  & -0.102   & 0.079   & -0.266** \\
& (0.156) & (0.148)  & (0.104) & (0.129)  \\
\# Days 19.5°C-21°C                       & 0.235   & -0.053   & -0.098  & 0.035    \\
& (0.143) & (0.152)  & (0.113) & (0.076)  \\
\# Days 21°C-22.5°C&     Ref.    &       Ref.     &   Ref.        &     Ref.       \\
&         &          &         &          \\
\# Days 22.5°C-24°C                       & 0.288** & -0.060   & 0.038   & 0.018    \\
& (0.129) & (0.128)  & (0.122) & (0.063)  \\
\# Days 24°C-25.5°C                       & 0.070   & 0.109    & 0.059   & 0.008    \\
& (0.119) & (0.132)  & (0.090) & (0.063)  \\
\# Days 25.5°C-27°C                       & 0.209   & 0.239    & 0.122   & 0.143    \\
& (0.138) & (0.157)  & (0.111) & (0.097)  \\
\# Days $\geq$ 27°C                       & 0.418*  & 0.677*** & 0.384** & 0.185    \\
& (0.237) & (0.238)  & (0.170) & (0.128)             \\
		&              &                     &            &                       \\
		Observations                                                 & 27,287       & 27,287              & 26,830     & 27,287                \\
		R-squared                                                    & 0.069        & 0.070               & 0.068      & 0.070                 \\
		Weather Controls  & \Checkmark              & \Checkmark                               & \Checkmark                 & \Checkmark                                    \\
		Year-month FE   & \Checkmark              & \Checkmark                               & \Checkmark                 & \Checkmark                                    \\
		State FE       & \Checkmark              & \Checkmark                               & \Checkmark                 & \Checkmark                                    \\
		KGC FE            & \Checkmark              & \Checkmark                               & \Checkmark                 & \Checkmark                                    \\
		\# of PSU                                                    & 371          & 371                 & 371        & 371                   \\
		Mean                                                         & 2.00         & 2.00                & 2.00       & 2.00                  \\ 
		\midrule
		\midrule
		\multicolumn{5}{p{20cm}}{\textit{Notes:} The outcome variable is binary and multiplied by 100, hence, coefficients are directly interpretable as percentage points. Robust standard  errors are clustered at the PSU level and presented in parentheses. ***   p\textless{}0.01, ** p\textless{}0.05, * p\textless{}0.1.}                  
	\end{tabular}
\end{table}
 \vspace*{\fill}
\end{landscape}

\setcounter{figure}{0} 
\renewcommand{\thefigure}{D\arabic{figure}}
\begin{figure}[p]
	\includegraphics[scale=.25]{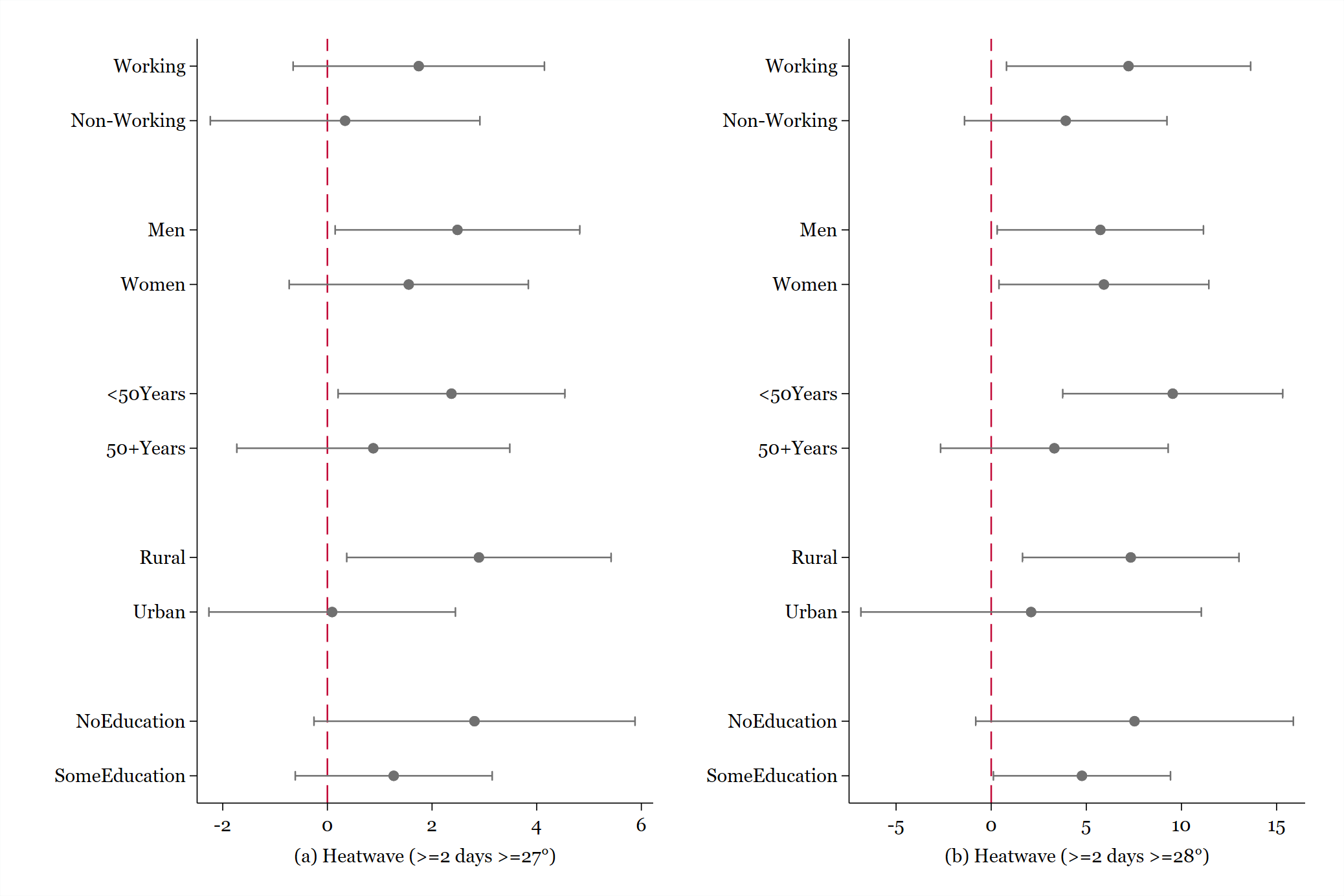}
	\centering
	\caption[Heterogeneous effects]{\textbf{Heterogeneous effects} - The figure displays the heat effect for different socioeconomic subgroups using a binary heatwave indicator. Graph (a) shows the effect for the binary heatwave indicator of two or more consecutive days above 27°C; graph (b) shows the effect for the binary heatwave indicator of two or more consecutive days above 28°C. The outcome variable is binary and multiplied by 100, hence, coefficients are directly interpretable as percentage points.}
	\label{hetero2}
\end{figure}

\clearpage
\subsection*{Appendix E: Roll-out of the District Mental Health Program}
\setcounter{table}{0} 
\renewcommand{\thetable}{E\arabic{table}}

\begin{table}[h!]
	\caption{\label{DMHP}Roll-out of the DMHP in the study districts}
	\centering
	\small
	\begin{tabular}{llll}
		\midrule
		\midrule
		(1) &	(2) &	(3 )&	(4) \\
State/UT	&	District	&	Year	&	WHO Sage Wave	\\
\midrule
Assam 	&	\textbf{Nagaon}	&	1996-97	&	Before Wave 0	\\
Assam 	&	\textbf{Goalpara}	&	1999-2000	&	Before Wave 0	\\
Assam 	&	\textbf{Darrang}	&	2004-05	&	Between Wave 0 \&1 	\\
Assam 	&	\textbf{Morigaon}	&	2004-05	&	Between Wave 0 \&1 	\\
Assam 	&	\textbf{Nalbari}	&	2004-05	&	Between Wave 0 \&1 	\\
Assam 	&	\textbf{Tinsukia}	&	2004-05	&	Between Wave 0 \&1 	\\
Karnataka	&	Chamrajnagar	&	2004-05	&		\\
Karnataka	&	\textbf{Gulbarga}	&	2004-05	&	Between Wave 0 \&1 	\\
Karnataka	&	Karwar	&	2004-05	&		\\
Karnataka	&	Shimoga	&	2004-05	&		\\
Karnataka	&	\textbf{Raichur}	&	2013-14	&	Between Wave 1 \& 2	\\
Karnataka	&	\textbf{Belgaum}	&	2013-14	&	Between Wave 1 \& 2	\\
Karnataka	&	\textbf{Dharwad}	&	2013-14	&	Between Wave 1 \& 2	\\
Karnataka	&	\textbf{Dakshin Kannada}	&	2013-14	&	Between Wave 1 \& 2	\\
Karnataka	&	Chikkaballapur	&	2013-14	&		\\
Karnataka	&	\textbf{Mysore}	&	2013-14	&	Between Wave 1 \& 2	\\
Karnataka	&	\textbf{Hassan}	&	2013-14	&	Between Wave 1 \& 2	\\
Karnataka	&	\textbf{Bellary}	&	2013-14	&	Between Wave 1 \& 2	\\
Maharashtra	&	\textbf{Raigard}	&	1997-98	&	Before Wave 0	\\
Maharashtra	&	Amravati	&	2003-04	&		\\
Maharashtra	&	\textbf{Buldhana}	&	2003-04	&	Between Wave 0\& 1 	\\
Maharashtra	&	\textbf{Parbhani}	&	2003-04	&	Between Wave 0 \& 1 	\\
Maharashtra	&	\textbf{Jalagaon}	&	2004-05	&	Between Wave 0 \& 1 	\\
Maharashtra	&	\textbf{Satara}	&	2004-05	&	Between Wave 0 \&1 	\\
Maharashtra	&	Alibag	&	2013-14	&		\\
Maharashtra	&	Nasik	&	2013-14	&		\\
Maharashtra	&	\textbf{Osmanabad}	&	2013-14	&	Between Wave 1 \& 2	\\
Maharashtra	&	Wrdha	&	2013-14	&	 \\
Maharashtra	&	\textbf{Bhandara}	&	2013-14	&	Between Wave 1 \& 2	\\
Maharashtra	&	Gadchiroli	&	2013-14	&		\\
Maharashtra	&	\textbf{Ahmednagar}	&	2013-14	&	Between Wave 1 \& 2	\\
Rajasthan	&	\textbf{Seekar}	&	1996-97	&	Before Wave 0	\\
Uttar Pradesh	&	\textbf{Kanpur}	&	1997-98	&	Before Wave 0	\\
Uttar Pradesh	&	Banda	&	2004-05	&		\\
Uttar Pradesh	&	\textbf{Faizabad}	&	2004-05	&	Between Wave 0 \&1 	\\
Uttar Pradesh	&	\textbf{Ghaziabad}	&	2004-05	&	Between Wave 0 \&1 	\\
Uttar Pradesh	&	\textbf{Itawah}	&	2004-05	&	Between Wave 0 \&1 	\\
Uttar Pradesh	&	\textbf{Mirzapur}	&	2004-05	&	Between Wave 0 \&1	\\
Uttar Pradesh	&	\textbf{Moradabad}	&	2004-05	&	Between Wave 0 \&1 	\\
Uttar Pradesh	&	Muzaffarnagar	&	2004-05	&		\\
Uttar Pradesh	&	\textbf{Raibareli}	&	2004-05	&	Between Wave 0 \&1 	\\
Uttar Pradesh	&	\textbf{Sitapur}	&	2004-05	&	Between Wave 0 \&1 	\\
West Bengal	&	\textbf{Bankura}	&	1998-99	&	Before Wave 0	\\
West Bengal	&	\textbf{Jalpaiguri}	&	2003-04	&	Between Wave 0 \&1 	\\
West Bengal	&	\textbf{West Midnapur}	&	2003-04	&	Between Wave 0 \&1 	\\
West Bengal	&	\textbf{South 24 Parganas}	&	2006-07	&	Between Wave 0 \&1 	\\
West Bengal	&	\textbf{Nadia}	&	2013-14	&	Between Wave 1 \& 2	\\
West Bengal	&	\textbf{Coochbehar}	&	2013-14	&	Between Wave 1 \& 2	\\

	\midrule
	\midrule
	\multicolumn{4}{p{11cm}}{\textit{Notes:} Districts in bold are those covered by the WHO-SAGE data. Information about the roll-out are extracted from \cite{dghs}.}
\end{tabular}
\end{table}

\clearpage

\end{document}